\documentclass[lettersize,journal]{IEEEtran}
\usepackage{amsmath,amsfonts}
\usepackage{algorithmic}
\usepackage{algorithm}
\usepackage{array}
\usepackage{textcomp}
\usepackage{stfloats}
\usepackage{url}
\usepackage{verbatim}
\usepackage{graphicx}
\graphicspath{{figures/}{./}}
\usepackage{subfigure}
\usepackage{subcaption}
\usepackage{cite}
\usepackage{color}
\usepackage{amssymb}
\usepackage{booktabs} 
\usepackage{caption}
\usepackage{balance}
\usepackage{geometry}
\usepackage{multirow}
\usepackage{hyperref}
\hypersetup{hidelinks,
	colorlinks=true,
	allcolors=black,
	pdfstartview=Fit,
	breaklinks=true,
	colorlinks=true,
	linkcolor=blue,
	citecolor=blue,
	urlcolor=blue}
\usepackage{amsmath,amsfonts,amsthm,bm} 
\usepackage[explicit]{titlesec}
\titlespacing{\subsection}{0pt}{1.5ex plus .0ex minus .0ex}{.3ex plus .0ex}

\titlespacing{\section}{0pt}{1.8ex plus .0ex minus .0ex}{1.5ex plus .0ex}
\makeatletter

\newcommand{\Rmnum}[1]{\expandafter\@slowromancap\romannumeral #1@}
\makeatother
\begin{document}

\title{Domain-Adaptive Deep Joint Source-Channel Coding for Image Classification}
\author{
\IEEEauthorblockN{Yishen Li, Xuechen~Chen,~\IEEEmembership{Member,~IEEE,}
	Xiaoheng~Deng,~\IEEEmembership{Senior~Member,~IEEE,}
	and~Hao~Zhang,~\IEEEmembership{Member,~IEEE}}
\thanks{This work was supported in part by Joint Funds for Railway Fundamental
	Research of National Natural Science Foundation of China under Grant
	No.~U2368201, by the National Natural Science Foundation of China under Grant Nos.~62172441, in part by Natural Science Foundation of Hunan Province under Grant No.2025JJ50388 and by the Changsha Natural Science Foundation under Grant kq2502118.
	Corresponding author: Xuechen Chen.
	
	Yishen Li, Xuechen Chen, Xiaoheng Deng and Hao Zhang are with the School of Electronic Information, Central South University, Changsha 410075, China (email: liyishen@csu.edu.cn; chenxuec@csu.edu.cn; dxh@csu.edu.cn; hao@csu.edu.cn).
}}

\maketitle
\pagestyle{empty}  
\thispagestyle{empty} 
\begin{abstract}
Deep joint source--channel coding (Deep JSCC) can suffer from domain shift between training and deployment data. We study single-source domain adaptation for image classification over wireless channels. A Gaussian example shows that improving channel quality does not necessarily improve target-domain accuracy under a fixed source-trained classifier. We further proposed an information objective to reduce domain information while preserving class information. Motivated by the analysis, we develop a domain-adaptive Deep JSCC method that combines class-wise adversarial alignment with pseudo-label supervised contrastive learning. The former uses all target samples to reduce class-conditional domain differences, while the latter selects high-confidence target samples to obtain class-discriminative received representations. We relate these losses to the information objective, and analyze the effects of pseudo-label errors and confidence-based selection. Experiments on SVHN--MNIST and PACS over AWGN and Rayleigh channels demonstrate improved target-domain classification without additional inference-time networks. The proposed method achieves 98.15\% accuracy on SVHN $\rightarrow$ MNIST at 10~dB CSNR. Code is available at \url{https://github.com/CSU-IPL/DAJSCC}.
\end{abstract}
\begin{IEEEkeywords}
visual semantic transmission, deep joint source-channel coding, domain adaptation, contrastive learning
\end{IEEEkeywords}

\section{Introduction}
With the rapid growth of intelligent visual applications, such as autonomous driving, industrial automation and mobile visual analytics, visual systems are increasingly required not only to capture and deliver data, but also to understand the semantic content of visual inputs for downstream tasks\cite{IOT1,IOT2}. In mobile and vehicular networks, this often involves transmitting visual data collected by onboard cameras and mobile devices over wireless links to edge servers or neighboring vehicles \cite{TVTcoinfer}. However, transmitting raw data is often inefficient, since a large portion of visual data is irrelevant to the target task \cite{SC1,TVTISCSC}. This has motivated growing interest in visual semantic understanding and transmission, where the goal is to extract task-relevant semantic information from images and convey compact visual representations that are sufficient for reliable inference at the receiver\cite{SC2,SC3,E2E,DRSW}. For example, in autonomous driving, a vehicle must be able to rapidly recognize key objects in its surroundings rather than reconstruct a full high-resolution image\cite{TVTvehicle,SC4}. In contrast, conventional communication transmits raw data such as images or video, which introduces non-negligible delay.

Deep Joint Source–Channel Coding (Deep JSCC) provides an effective end-to-end framework for visual semantic understanding and transmission\cite{djscc1, djscc2, djscc3}. Deep JSCC employs deep neural networks to learn a direct mapping from source signals to channel symbols, and it maps the received symbols directly to task outputs at the receiver. Unlike traditional separated schemes, Deep JSCC enables end-to-end joint optimization of source compression, channel coding and task inference. This transmission paradigm reduces the delivery of task-irrelevant information and can lower the inference latency of intelligent visual systems. It has been shown to be effective in various downstream tasks, such as image classification~\cite{image}, speech recognition~\cite{speech}, and video analysis~\cite{video,zhu2023}. Recent studies have further developed task-oriented Deep JSCC frameworks for text and visual data, aiming to learn compact semantic representations that preserve task-relevant information while suppressing redundant content~\cite{task1,task2,task3}. In addition, information-bottleneck-based methods reduce communication cost by sparsifying the encoded representation and removing irrelevant dimensions~\cite{task4}.

Despite these advances, domain shift remains a challenge for Deep JSCC-based visual semantic transmission. Deployment images may follow a different distribution from the training data, degrading target-domain classification
performance~\cite{OOD1}. Several studies have addressed this problem in Deep JSCC. Zhang et al.~\cite{OOD2} introduce a data adaptation module before semantic encoding, using a generative adversarial network (GAN)~\cite{GAN} to transform inputs toward the training distribution. Won et al.~\cite{OOD4} study multi-source domain
adaptation using StarGAN-based domain translation. However, both of them require an additional generative network during deployment, which leads to extra computational overhead. Li et al.~\cite{OOD3} combine the information
bottleneck principle with invariant risk minimization to improve generalization across training environments. It focuses on multi-domain generalization and does not consider the problem of single-domain adaptation.
We aim to adapt the end-to-end JSCC system for image classification task in a single-source unsupervised domain adaptation setting, without the need for additional generative modules.

Conventional domain adaptation provides useful tools for learning transferable representations. MMD-based methods match source and target feature distributions~\cite{MMD1,MMD2}, DANN uses adversarial domain discrimination~\cite{DANN,ma2019}, and CORAL aligns feature covariance matrices~\cite{CORAL}. Related efforts explore uncertainty-aware unsupervised adaptation~\cite{guan2022} and joint distribution matching~\cite{KMUR}.
These studies highlight the importance of representations that are both transferable and discriminative; however, reducing domain discrepancy alone does not guarantee low error on the target domain~\cite{zhao2019invariant}.
These methods generally do not account for how wireless transmission affects domain adaptation at the receiver. Moreover, the effects of pseudo-label errors and confidence-based sample selection on the received information objective remain unclear. In our setting, the learned transmitter and physical channel jointly determine the received
representations used for classification. Channel noise can suppress conditional domain information while also removing class information, so a smaller received domain discrepancy does not necessarily imply better classification. Therefore, we aim to reduce class-conditional differences across domains while preserving class information in the received representations. This motivates joint consideration of domain alignment and class discriminability at the receiver.

In this paper, we study single-source unsupervised domain adaptation for image classification over wireless channels. We analyze conditional domain information $I(D;\widehat Z\mid Y)$ and class information $I(Y;\widehat Z)$ at the channel output. This analysis motivates a joint design that reduces class-conditional domain differences while preserving class-discriminative received
representations. We implement these objectives through class-wise adversarial alignment and pseudo-label contrastive learning. The alignment uses source labels and target pseudo labels, while contrastive learning organizes labeled source and high-confidence target samples by class. The adaptation modules are used only during training, without requiring additional inference modules. The main contributions are summarized as follows.
\begin{itemize}
\item[$\bullet$] We propose a domain-adaptive Deep JSCC framework for wireless image classification, which is trained using a labeled source domain and an unlabeled target domain. We formulate a received-signal information objective that combines source classification, class-conditional domain alignment, and class-information preservation to learn transferable and discriminative representations under wireless transmission.

\item[$\bullet$] We design a received-signal adaptation method combining class-wise adversarial alignment and confidence-filtered pseudo-label contrastive learning. We connect the optimized adversarial loss to conditional domain information and establish a class-information lower bound associated with the contrastive loss. The method jointly trains the JSCC system without additional inference modules.

\item[$\bullet$] We analyze how target pseudo-label errors and confidence filtering affect the information objective. We derive an upper bound relating the ideal and pseudo-label information objectives, while accounting for pseudo-label errors in conditional alignment and contrastive learning, as well as filtering-induced selection bias. This analysis characterizes the tradeoff between pseudo-label reliability and sample coverage.

\item[$\bullet$] We evaluate the proposed method on SVHN$\rightarrow$MNIST and six PACS adaptation tasks over AWGN and Rayleigh fading channels. Comparisons with baseline methods demonstrate improved target-domain classification. Ablation studies further clarify the contributions of the proposed components. On SVHN$\rightarrow$MNIST, the method achieves $98.15\%$ target accuracy over an AWGN channel at a CSNR of $10$~dB.
\end{itemize}

The remainder of this paper is organized as follows. In Section II, the problem is stated. Section III presents the channel-aware information analysis. The proposed Deep JSCC method and its implementation are described in Section IV. Evaluations of the proposed method are reported in Section V. Finally, the conclusions are presented in Section VI.

$Notation$: Bold uppercase and lowercase letters represent random vectors and their realizations, respectively. Uppercase and lowercase letters represent random variables and their realizations, respectively. 
\begin{figure}[!t]
\setlength{\belowcaptionskip}{-0.3cm} 
\vspace{-0.2cm}
\centering
\includegraphics[width=3.3in,height=1.4in]{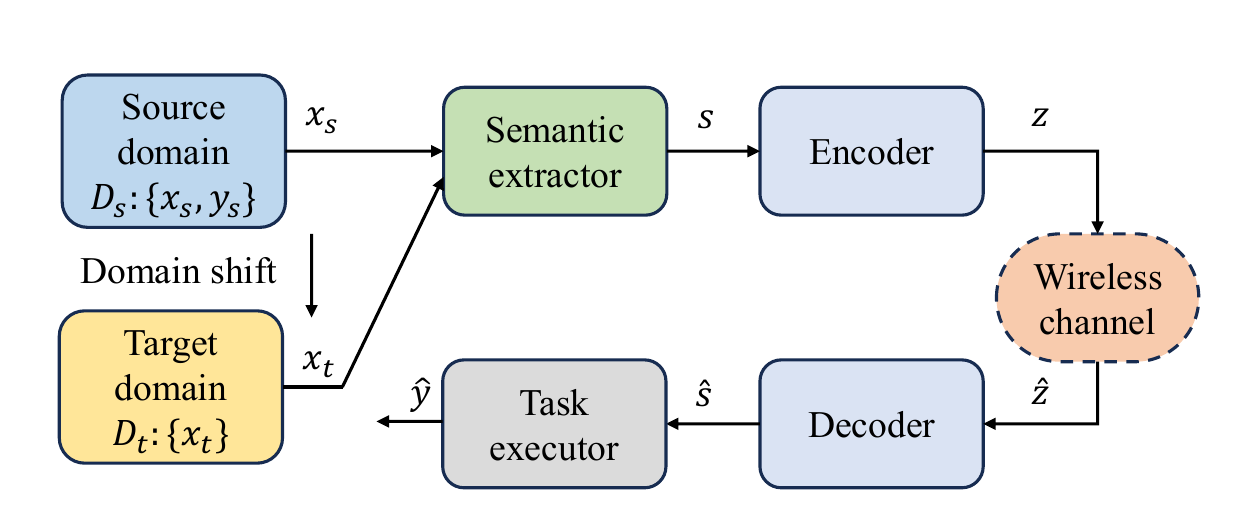}
\caption{Overview of the task-oriented deep JSCC system.}
\label{OW}
\end{figure}
\section{Problem Statement}
\subsection{System Model}
In this paper, we study a task-oriented deep JSCC system, as illustrated in Fig. \ref{OW}. The system is considered under domain shift, where the data distribution at deployment differs from that during training, referred to as the target domain and the source domain, respectively. During training, the system has access to labeled data from the source domain and a subset of unlabeled samples from the target domain to facilitate adaptation. Specifically, we have a source domain $D_s$ with $n_s$ labeled samples and a target domain $D_t$ with $n_t$ unlabeled samples. Let $\mathcal{X}_s$ and $\mathcal{X}_t$ be the source and target sample spaces, with individual samples $\mathbf{x}_s \in \mathcal{X}_s$ and $\mathbf{x}_t \in \mathcal{X}_t$. Define $\mathcal{X} = \mathcal{X}_s \cup \mathcal{X}_t$ and use $\mathbf{x} \in \mathcal{X}$ to denote a generic sample from either domain. The semantic extractor $f_{\theta_{f}}: \mathbb{R}^{n} \rightarrow \mathbb{R}^{l}$ maps input  $\mathbf{x}$ to a semantic representation $\mathbf{s}$. The encoder $f_{\theta_e}: \mathbb{R}^{l} \rightarrow \mathbb{C}^{m}$ maps $\mathbf{s}$ to a complex representation $\mathbf{z}$; $f_{\theta_f}$ and $f_{\theta_e}$ are parameterized by neural networks with parameters $\theta_f$ and $\theta_e$. Let $\hat{\mathbf{z}}$ denote the received symbol corresponding to $\mathbf{z}$ after transmission over a noisy channel. We model the channel as
\begin{align}
	\hat{\mathbf{z}} = \mathbf{h} \odot \mathbf{z} + \mathbf{n},
\label{channel}
\end{align}
where $\mathbf{h}\sim\mathcal{CN}(\mathbf{0},\mathbf{I})$ is the channel-gain vector for Rayleigh fading, $\mathbf{n}\sim\mathcal{CN}(\mathbf{0},\sigma_n^2\mathbf{I})$ is additive white Gaussian noise, and $\odot$ denotes component-wise multiplication. The AWGN channel is obtained as the special case $\mathbf{h}=\mathbf{1}$.

To meet the practical energy constraints, we impose an average power limit $P$ on the encoder output. Specifically, with $\mathbf{z} = f_{\theta_e}(\mathbf{s})$, we require $\frac{1}{m}\mathbb{E}[\|\mathbf{z}\|^{2}_{2}]\leq P$. The channel quality can be measured by Channel Signal-to-Noise Ratio (CSNR),
\begin{align}
	\rm{CSNR} = 10\log_{10}\left(\frac{P}{\sigma^2_n}\right) \rm{dB}.
\end{align}

At the receiver, the decoder $f_{\phi_d}(\cdot)$ recovers the semantic feature $\hat{\mathbf{s}}=f_{\phi_d}(\hat{\mathbf{z}})$ from the received signal. The task executor $f_{\phi_t}(\cdot)$ then produces the predicted label $\hat y=f_{\phi_t}(\hat{\mathbf{s}})$. Since source samples are labeled and target samples are unlabeled, the supervised classification loss is evaluated on the source domain, while target samples contribute to unsupervised adaptation.

We seek to improve target-domain classification by learning received representations with similar class-conditional distributions across domains and clear separation between classes. Specifically, source and target features associated with class $k$ should have similar distributions, $P_{\widehat Z\mid D=s,Y=k}$ and $P_{\widehat Z\mid D=t,Y=k}$. The transmitted features are affected by the wireless channel before classification, so the alignment and discriminability of the received features are central to the design. We jointly optimize the semantic extractor, encoder, decoder, and classifier under the communication power and bandwidth constraints.

\section{Channel-Aware Information Analysis}
\label{sec:channel_theory}
In this section, we first show that classification accuracy in the target domain can decrease as channel quality improves using a Gaussian example. We then relate the class-conditional domain information to target classification risk, and introduce a complementary class-information objective for adaptation with unlabeled target data.

\begin{figure*}[!t]
	\centering
	\begin{minipage}[t]{0.47\textwidth}
		\centering
		\includegraphics[width=3.0in,height=2.0in]{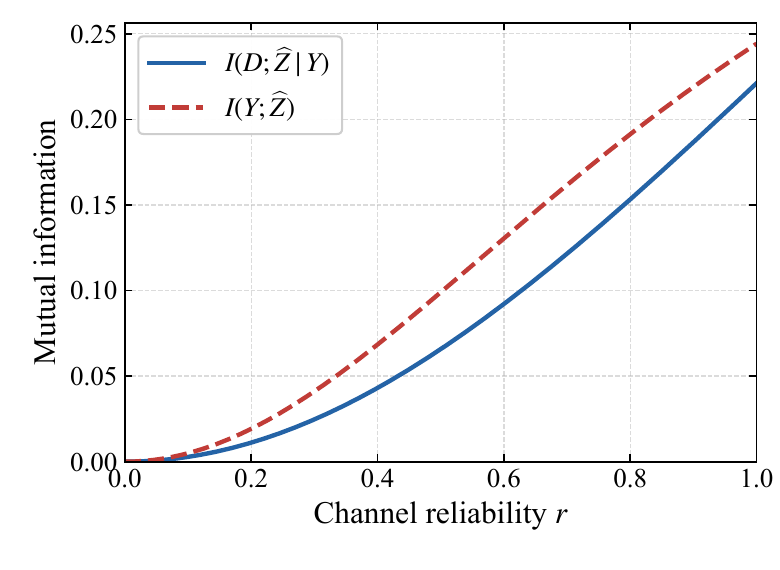}
		\par\vspace{3pt}
		{\footnotesize (a) Received domain and class information.\par}
	\end{minipage}
	\hfill
	\begin{minipage}[t]{0.47\textwidth}
		\centering
		\includegraphics[width=3.0in,height=2.0in]{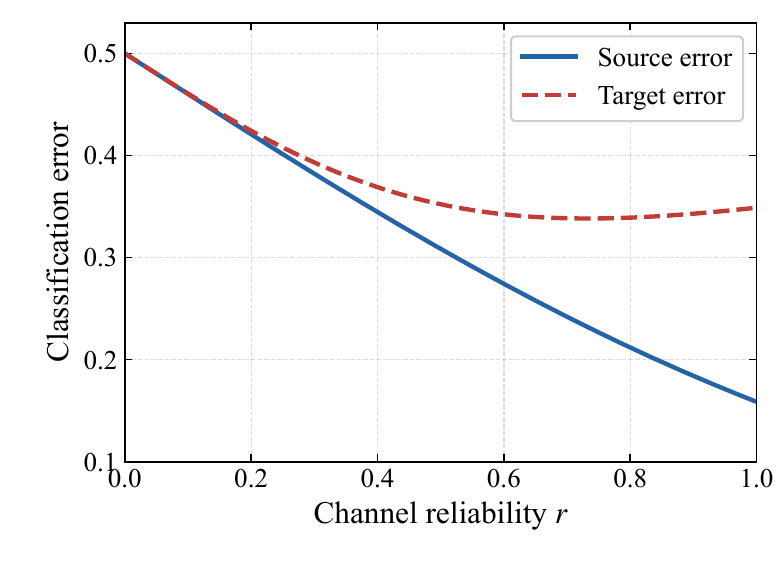}
		\par\vspace{3pt}
		{\footnotesize (b) Source and target classification errors.\par}
	\end{minipage}
	\caption{Gaussian information and classification error versus channel reliability $r$.}
	\label{fig:gaussian_information_error}
\end{figure*}

\subsection{Gaussian Example Under Domain Shift}
Let $D\in\{s,t\}$ denote the source or target domain, and let $Y$ denote the class label. Consider a binary Gaussian example to examine how channel reliability affects received information and cross-domain classification. We set two classes which have equal priors and
\begin{equation}
	\begin{aligned}
		X_s\mid Y=1&\sim\mathcal N(-1,1), &
		X_s\mid Y=2&\sim\mathcal N(1,1),\\
		X_t\mid Y=1&\sim\mathcal N(0.5,1), &
		X_t\mid Y=2&\sim\mathcal N(2.5,1).\notag
	\end{aligned}
	\label{eq:gaussian_domains}
\end{equation}

Let $\widehat Z=eX+N$, where $e>0$ and
$N\sim\mathcal N(0,\sigma_n^2)$ is independent of $X$,
$D$, and $Y$. The source-optimal decision threshold
is zero at every noise level. Defining
$r=e/\sqrt{e^2+\sigma_n^2}\in(0,1]$, which represents the channel reliability. The source and target classification errors are
\begin{equation}
	\begin{aligned}
		R_s(r)=\Phi(-r),
		R_t(r)=\tfrac12[\Phi(0.5r)+\Phi(-2.5r)],
	\end{aligned}
	\label{eq:gaussian_target_risk}
\end{equation}
where $\Phi$ is the standard normal CDF.
The target-error derivative is
\begin{equation}
	R_t'(r)=\tfrac14\varphi(0.5r)
	\big[1-5\exp(-3r^2)\big],\notag
	\label{eq:gaussian_target_derivative}
\end{equation}
where $\varphi$ is the standard normal density.
Hence, source error decreases monotonically as $r$ increases, whereas target error first decreases and then increases, attaining its minimum at $r=\sqrt{\ln 5/3}$. The received class-conditional distributions are
\begin{equation}
	\widehat Z\mid D=d,Y=k
	\sim\mathcal N(e\mu_{d,k},e^2+\sigma_n^2).\notag
	\label{eq:gaussian_received_distribution}
\end{equation}
Scaling it by $\sqrt{e^2+\sigma_n^2}$ gives means
$r\mu_{d,k}$ and unit variance without changing mutual
information. We therefore evaluate
$I(D;\widehat Z\!\mid\! Y)$ and $I(Y;\widehat Z)$ as functions
of $r$ by one-dimensional numerical integration of
the Gaussian mixture densities.

Fig.~\ref{fig:gaussian_information_error} shows that
both $I(D;\widehat Z\!\mid\! Y)$ and $I(Y;\widehat Z)$
increase with channel reliability.
Channel noise suppresses domain information but also
removes class information, so a smaller conditional
domain discrepancy alone does not imply better
classification. Higher channel reliability provides
more class information but also brings more domain-unique information, resulting in a decrease in target domain accuracy.

\subsection{Class-Conditional Domain Information}
For the general $K$-class analysis, assume equal class priors in the source and target domains,
\begin{equation}
	\Pr(Y=k\mid D=s)=\Pr(Y=k\mid D=t)=\pi_k,
	\label{eq:common_prior}
\end{equation}
and set $\Pr(D=s)=\Pr(D=t)=1/2$. Let $Z$ and $\widehat Z$ denote transmitted and received representations, and write $P_{d,k}=P_{\widehat Z\mid D=d,Y=k}$. The decoder and classifier define a shared decision rule $g_\phi(\widehat Z)$, with transmitter parameters $\theta=(\theta_f,\theta_e)$ and receiver parameters $\phi=(\phi_d,\phi_t)$. For $d\in\{s,t\}$, define the classification risk as $R_d=\Pr[g_\phi(\widehat Z)\ne Y\mid D=d]$.

The conditional mutual information $I(D;\widehat Z\mid Y)$ measures how distinguishable the two domains remain within each true class at the receiver. With $M_k=(P_{s,k}+P_{t,k})/2$, we have
\begin{equation}
	\begin{aligned}
		I(D;\widehat Z\mid Y=k)
		={}&\tfrac12 D_{\mathrm{KL}}(P_{s,k}\Vert M_k)
		+\tfrac12 D_{\mathrm{KL}}(P_{t,k}\Vert M_k)\\
		={}&D_{\mathrm{JS}}(P_{s,k}\Vert P_{t,k}),
	\end{aligned}
	\label{eq:class_conditional_js}
\end{equation}
where $D_{\mathrm{JS}}$ denotes the Jensen--Shannon divergence. Averaging across classes gives
\begin{equation}
	I(D;\widehat Z\mid Y)
	=\sum_{k=1}^{K}\pi_k D_{\mathrm{JS}}(P_{s,k}\Vert P_{t,k}).
	\label{eq:conditional_domain_information}
\end{equation}

The connection to target-domain performance can be made precise. For a shared classifier $g_\phi$, define
$A_k=\{\hat z:g_\phi(\hat z)\ne k\}$.
Under the common class priors in
\eqref{eq:common_prior}, the risk difference
satisfies
\begin{equation}
	\begin{aligned}
		|R_t-R_s|
		&\le \sum_{k=1}^{K}\pi_k
		\operatorname{TV}(P_{s,k},P_{t,k})\\
		&\le \sum_{k=1}^{K}\pi_k
		\sqrt{2D_{\mathrm{JS}}(P_{s,k}\Vert P_{t,k})}\\
		&\le \sqrt{2\sum_{k=1}^{K}\pi_k
			D_{\mathrm{JS}}(P_{s,k}\Vert P_{t,k})}\\
		&=\sqrt{2I(D;\widehat Z\mid Y)}.
	\end{aligned}
	\label{eq:target_risk_bound}
\end{equation}
The first inequality follows from the triangle
inequality and the definition of total variation.
The second follows from Pinsker's inequality,
while the third uses Jensen's inequality.
The last equality follows from
\eqref{eq:conditional_domain_information}.
This bound shows that reducing the received domain discrepancy controls the target risk relative to the source risk, which motivates jointly minimizing source risk
and received class-conditional domain information,
\begin{equation}
	\min_{\theta,\phi}\quad
	R_s+\lambda I(D;\widehat Z\mid Y),
	\label{eq:initial_alignment_objective}
\end{equation}
where $\lambda>0$ controls the alignment weight.

For a fixed channel configuration $\eta$
with transition law $Q_\eta$, the received
distributions satisfy
\begin{equation}
	P_{\widehat Z\mid D=d,Y=k}
	=Q_\eta P_{Z\mid D=d,Y=k}.
	\label{eq:channel_distribution_mapping}
\end{equation}
Thus, the distributions in the risk bound
depend on both the encoder and the physical channel. We therefore apply alignment
after transmission to directly reduce the
class-conditional discrepancy in the representations
used by the receiver for classification.

\subsection{Joint Design Objective}
In practical unsupervised adaptation, target-domain
labels are unavailable. Source classification
provides direct label supervision only for source
samples, whereas the target domain lacks explicit supervision. Conditional alignment reduces discrepancies between corresponding classes across
domains, but does not itself enforce separation between
different classes. Besides, the Gaussian example shows
that channel noise can suppress class information along
with domain information. These observations motivate
explicitly encouraging class-discriminative received
representations across both domains.

To provide additional class-discriminative supervision across both domains, we introduce $I(Y;\widehat Z)=H(Y)-H(Y\mid\widehat Z)$. This term can be maximized to reduce class uncertainty and encourage class-discriminative received representations.
Combining this regularizer with source classification
and conditional alignment leads to
\begin{equation}
	\mathcal L_{\mathrm{ideal}}
	=R_s+\lambda I(D;\widehat Z\mid Y)
	-\beta I(Y;\widehat Z),
	\label{eq:information_design_objective}
\end{equation}
where $\lambda,\beta>0$ control alignment and
class-information regularization, respectively. Note that direct evaluation of these mutual-information terms is generally intractable for the learned high-dimensional
received distributions. Section~IV therefore develops
trainable losses for conditional alignment
and class-information terms.

\begin{figure*}[!t]
	\setlength{\belowcaptionskip}{-0.3cm} 
	\vspace{-0.2cm}
	\centering
	\includegraphics[width=6.4in,height=3.2in]{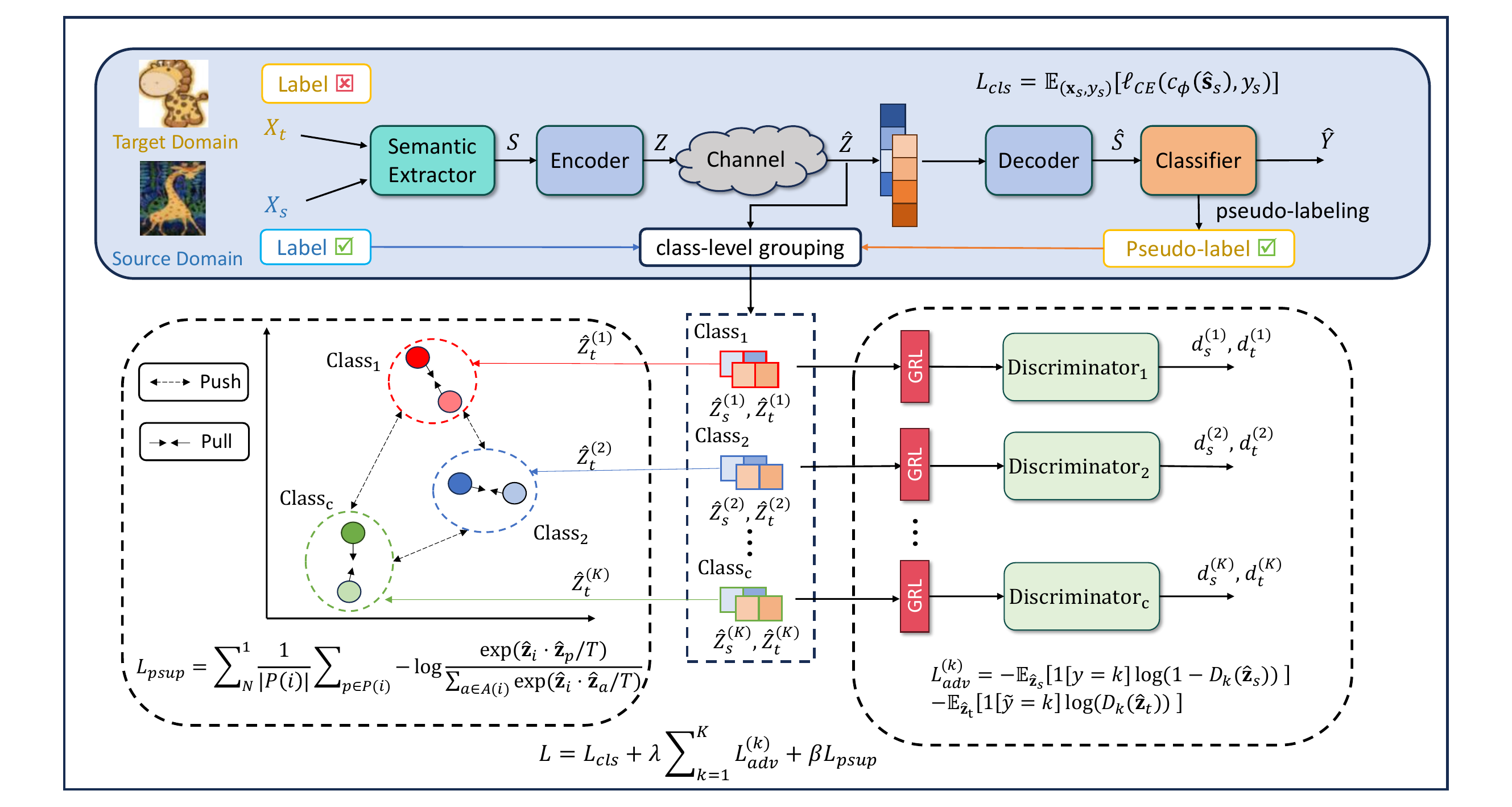}
	\caption{Overview of the proposed domain-adaptive joint source-channel coding scheme with class-level alignment and pseudo-label supervised contrastive learning.}
	\label{DAJSCC}
\end{figure*}
\section{Proposed Deep JSCC Method}
Motivated by Section~\ref{sec:channel_theory}, we combine
class-wise adversarial alignment with contrastive
learning on received representations, as shown in
Fig.~\ref{DAJSCC}. Besides, we investigated the connection between the training loss and the proposed optimization objective in \eqref{eq:information_design_objective}.

\subsection{Class-level Domain Alignment}
The transmitter produces $\mathbf z=f_{\theta_e}(f_{\theta_f}(\mathbf x))$, and the receiver predicts the class through $g_\phi(\hat{\mathbf z})=f_{\phi_t}(f_{\phi_d}(\hat{\mathbf z}))$. We train the source classifier with cross-entropy loss,
\begin{equation}
\mathcal L_{\mathrm{cls}}=
\mathbb E\big[\ell_{\mathrm{CE}}(g_\phi(\widehat Z_s),Y_s)\big],
\label{eq:method_source_ce}
\end{equation}

The initial design criterion in \eqref{eq:initial_alignment_objective} motivates reducing $I(D;\widehat Z\mid Y)$. Equation~\eqref{eq:conditional_domain_information} expresses this quantity as a weighted sum of class-conditional Jensen--Shannon (JS) divergences. We therefore introduce domain discriminator $D_k$ for each class and train the discriminators using received features instead of aligning only the marginal source and target distributions.

We first analyze such alignment using true labels in both domains. For the theoretical analysis, we define a source--target mixture with $P(D=s)=P(D=t)=1/2$ and common class priors $\pi_k$.  Equal domain weighting treats the two domains symmetrically, while the common-prior assumption allows us to focus on class-conditional distribution shifts. Let $p_{d,k}$ denote the density of
$P_{d,k}$. After the transmission, we can obtain
\begin{equation}
	p_{d,k}(\hat{\mathbf z})
	=\int q_\eta(\hat{\mathbf z}\mid\mathbf z)\,
	p(\mathbf z\mid D=d,Y=k)\,
	\mathrm d\mathbf z.
	\label{eq:channel_induced_density}
\end{equation}

For class $k$, we consider the domain discrimination objective
\begin{equation}
	\begin{aligned}
		V_k(D_k)
		={}&\mathbb E_{P_{s,k}}
		\big[\ln(1-D_k(\widehat Z))\big]\\
		&+\mathbb E_{P_{t,k}}
		\big[\ln D_k(\widehat Z)\big],
	\end{aligned}
	\label{eq:ideal_class_discriminator}
\end{equation}
where $D_k(\hat{\mathbf z})$ is the predicted
probability that a received sample comes from
the target domain. Following the standard adversarial discrimination
analysis \cite{2014GAN}, the optimal discriminator
for fixed received distributions is
\begin{equation}
	D_k^*(\hat{\mathbf z})
	=\frac{p_{t,k}(\hat{\mathbf z})}
	{p_{s,k}(\hat{\mathbf z})
		+p_{t,k}(\hat{\mathbf z})}.
	\label{eq:ideal_domain_discriminator_opt}
\end{equation}
Substituting the optimal discriminators into
\eqref{eq:ideal_class_discriminator} gives
\begin{equation}
	\begin{aligned}
		\max_{\{D_k\}}\sum_{k=1}^{K}\pi_kV_k(D_k)
		&=2\mathbb E[\ln P(D\mid\widehat Z,Y)]\\
		&=-2H(D\mid\widehat Z,Y)\\
		&=-2[H(D)-I(D;\widehat Z,Y)]\\
		&=-\ln4+2I(D;\widehat Z,Y)\\
		&=-\ln4+2I(D;\widehat Z\mid Y).
	\end{aligned}
	\label{eq:class_adversarial_information}
\end{equation}
Here, $H(D)=\ln 2$ follows from the equally weighted source-target mixture, and the last equality holds because the common class priors imply $I(D;Y)=0$. Consequently, with optimal discriminators, minimizing the discrimination objective over the transmitter is equivalent to minimizing $I(D;\widehat Z\!\mid\! Y)$. This motivates the use of class-wise discrimination on received features.

In practice, we use pseudo labels for target samples. Let $\hat p_{i,k}$ be the classifier's predicted probability for class $k$,
\begin{equation}
\tilde y_i=\arg\max_{1\le k\le K}\hat p_{i,k},\qquad
q_i=\max_{1\le k\le K}\hat p_{i,k}.
\label{eq:target_pseudo_confidence}
\end{equation}
Source samples use the ground-truth labels, whereas target samples use $\tilde y_i$. We denote the label used during adaptation as
\begin{equation}
\bar y_i=\begin{cases}
 y_i,&\mathbf x_i\in D_s,\\
 \tilde y_i,&\mathbf x_i\in D_t.
\end{cases}
\label{eq:adaptation_label}
\end{equation}

The class-$k$ domain-discrimination loss is
\begin{equation}
\begin{aligned}
\mathcal L_{\mathrm{adv}}^{(k)}={}&-\mathbb E_{s}\big[\mathbf1(Y_s=k)
\ln(1-D_k(\widehat Z_s))\big]\\
&-\mathbb E_{t}\big[\mathbf1(\widetilde Y_t=k)
\ln D_k(\widehat Z_t)\big],
\end{aligned}
\label{eq:actual_class_adv}
\end{equation}

The overall adversarial loss is
\begin{equation}
\mathcal L_{\mathrm{adv}}=\sum_{k=1}^{K}\mathcal L_{\mathrm{adv}}^{(k)}.
\label{DALoss}
\end{equation}

\begin{figure}[t]
	\setlength{\belowcaptionskip}{-0.3cm} 
	\vspace{-0.4cm}
	\centering
	\includegraphics[width=2.5in,height=1.2in]{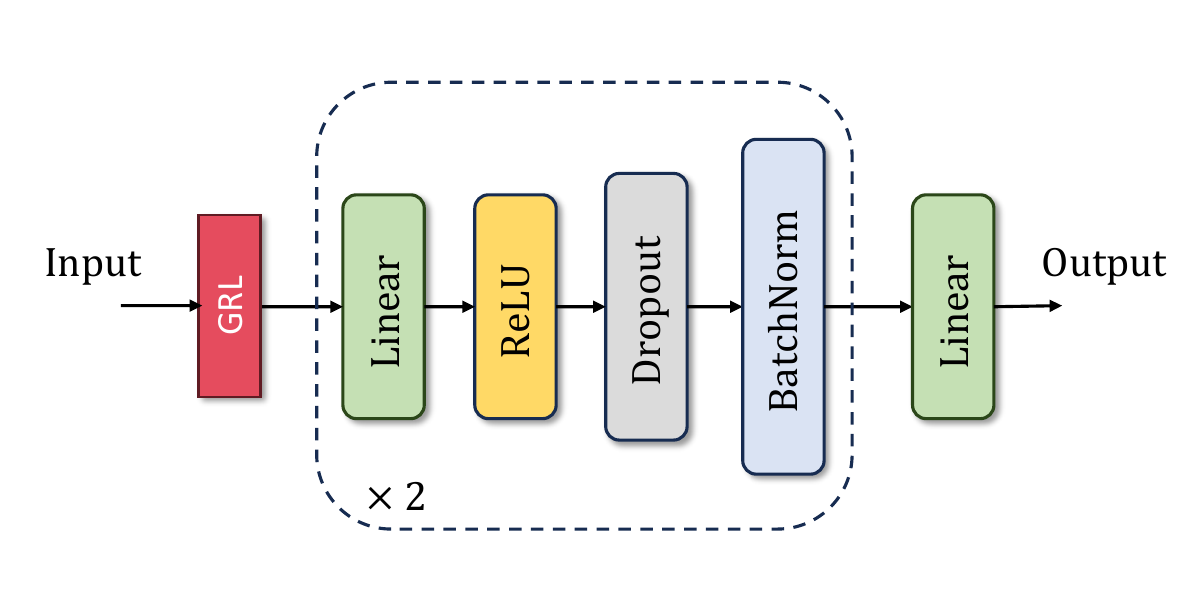}
	\caption{The specific structure of the discriminator.}
	\label{Dis}
\end{figure}

The domain discriminators minimize $\mathcal L_{\mathrm{adv}}$ to identify the two domains, while a gradient reversal layer (GRL) reverses the gradient reaching the transmitter to make the received features less domain dependent. The GRL acts as the identity in the forward pass and changes the sign of the gradient in the backward pass. The contribution of this loss to the transmitter gradient is $-\lambda\nabla_{\theta}\mathcal L_{\mathrm{adv}}$, where $\theta=(\theta_f,\theta_e)$. Thus the practical optimization follows the minimax direction suggested by \eqref{eq:class_adversarial_information}, using pseudo-label-defined class groups. Fig.~\ref{Dis} shows the discriminator architecture, which contains linear layers with 128 and 64 units followed by a sigmoid output.

\subsection{Pseudo-label Contrastive Learning}
Class-wise adversarial learning reduces class-conditional discrepancies between source and target received representations. To further encourage class discriminability in the source–target mixture, we introduce pseudo-label contrastive learning (PCL) as an auxiliary regularization. Specifically, PCL exploits ground-truth source labels and high-confidence target predictions to construct positive and negative sample pairs, promoting within-class similarity and between-class separation in the received feature space.

We next establish a connection between supervised contrastive learning and the class information $I(Y;\widehat Z)$ under ideal class-label supervision.  Consider an anchor $(Y_i,\widehat Z_i)$ drawn from the source--target mixture. Let $\mathcal A$ be a reference set independent of the anchor and it contains $M\ge1$ samples from each class under the same channel configuration. The positive set $\mathcal P(i)$ contains the $M$ reference samples of class $Y_i$. The similarity of the anchor to reference sample $a$ is $s_{ia}=\operatorname{sim}(\hat{\mathbf z}_i,\hat{\mathbf z}_a)/{T}$,  where $\operatorname{sim}(\hat{\mathbf z}_i,\hat{\mathbf z}_a)
=(\hat{\mathbf z}_i^{\mathsf T}\hat{\mathbf z}_a)/
(\|\hat{\mathbf z}_i\|_2\|\hat{\mathbf z}_a\|_2)$. The supervised contrastive loss \cite{SupLoss} is
\begin{equation}
\ell_i^{\mathrm{sup}}=\ln\!\sum_{a\in\mathcal A}e^{s_{ia}}
-\frac1M\sum_{p\in\mathcal P(i)}s_{ip},
\label{eq:ideal_supcon_loss}
\end{equation}
where $T>0$ is a temperature and $\operatorname{sim}(\hat{\mathbf z}_i,\hat{\mathbf z}_a)$ measures similarity between two received features.

To relate this loss to class information in our objective, we first normalize the reference similarities into probabilities $w_{ia}=e^{s_{ia}}/\sum_{b\in\mathcal A}e^{s_{ib}}$. After aggregating these probabilities by reference class, we can get a class predictor
\begin{equation}
q_{\mathcal A}(y\mid\hat{\mathbf z}_i)
=\sum_{a\in\mathcal A:\,y_a=y}w_{ia}
=\frac{\sum_{a\in\mathcal A:\,y_a=y}e^{s_{ia}}}
{\sum_{a\in\mathcal A}e^{s_{ia}}}.
\label{eq:contrastive_class_posterior}
\end{equation}

Let $\mathcal L_{\mathrm{cc}}=\mathbb E[-\ln q_{\mathcal A}(Y_i\mid\widehat Z_i)]$ be the expected cross-entropy of this predictor. Since $\mathcal A$ is independent of the anchor, the cross-entropy decomposition gives
\begin{equation}
\begin{aligned}
\mathcal L_{\mathrm{cc}}
&=H(Y\mid\widehat Z)+\mathbb E_{\mathcal A,\widehat Z}
D_{\mathrm{KL}}\!\left(P_{Y\mid\widehat Z}\Vert
q_{\mathcal A}(Y\mid\widehat Z)\right)\\
&\ge H(Y\mid\widehat Z).
\end{aligned}
\label{eq:class_ce_bound}
\end{equation}
Thus $I(Y;\widehat Z)=H(Y)-H(Y\!\mid\!\widehat Z)\ge H(Y)-\mathcal L_{\mathrm{cc}}$. To connect $\mathcal L_{\mathrm{cc}}$ to the contrastive loss for each anchor, use Jensen's inequality, we get
\begin{equation}
\ln\!\sum_{p\in\mathcal P(i)}e^{s_{ip}}
\ge \ln M+\frac1M\sum_{p\in\mathcal P(i)}s_{ip}.
\label{eq:positive_jensen}
\end{equation}

Using \eqref{eq:contrastive_class_posterior} and \eqref{eq:positive_jensen}, the class cross-entropy for that anchor satisfies
\begin{equation}
-\ln q_{\mathcal A}(Y_i\mid\hat{\mathbf z}_i)
\le\ell_i^{\mathrm{sup}}-\ln M.
\label{eq:contrastive_ce_comparison}
\end{equation}

Taking expectations and applying \eqref{eq:class_ce_bound} gives
\begin{equation}
I(Y;\widehat Z)\ge H(Y)+\ln M-
\mathcal L_{\mathrm{sup}},
\label{eq:pcl_information_lower_bound}
\end{equation}
where $\mathcal L_{\mathrm{sup}}=\mathbb E[\ell_i^{\mathrm{sup}}]$. Minimizing the supervised contrastive loss therefore increases a lower bound on the class information of the source-target mixture.

In the actual implementation, target labels are unavailable. We use all labeled source samples and only target samples whose confidence in \eqref{eq:target_pseudo_confidence} exceeds $\tau$. The indices available for contrastive learning are
\begin{equation}
\mathcal I=\{i:\mathbf x_i\in D_s\}\cup
\{i:\mathbf x_i\in D_t,\ q_i>\tau\}.\notag
\label{indice}
\end{equation}

For an anchor $i\in\mathcal I$, let
$\mathcal A(i)=\mathcal I\setminus\{i\}$ and
$\mathcal P(i)=\{j\in\mathcal A(i):\bar y_j=\bar y_i\}$,
with $\bar y_i$ defined by \eqref{eq:adaptation_label}. Anchors without a positive sample are omitted, so $\mathcal I_+=\{i\in\mathcal I:|\mathcal P(i)|>0\}$. The contrastive form in \eqref{eq:ideal_supcon_loss} becomes
\begin{equation}
\begin{aligned}
\ell_i^{\mathrm{psup}}
&=-\frac1{|\mathcal P(i)|}\sum_{p\in\mathcal P(i)}
\ln\frac{e^{s_{ip}}}{\sum_{a\in\mathcal A(i)}e^{s_{ia}}},\\
\mathcal L_{\mathrm{psup}}
&=\frac1{|\mathcal I_+|}\sum_{i\in\mathcal I_+}\ell_i^{\mathrm{psup}}.
\end{aligned}
\label{ConLoss}
\end{equation}
This loss is evaluated on channel-impaired features, encouraging within-class similarity and between-class separation in the representations available to the receiver. 

\subsection{Analysis of Pseudo-Label Errors}
Since target-domain pseudo-labels are used during adaptation, we next analyze how their errors affect the information objective. CDA uses all target samples to cover the target-domain distribution. In contrast, PCL uses only high-confidence samples because contrastive learning requires more reliable positive samples \cite{CRST}. We therefore account for their respective pseudo-label errors and the selection bias introduced by PCL.

Let $\bar Y$ denote the label used during adaptation, as defined in \eqref{eq:adaptation_label}. Following the same optimal-discriminator analysis as in Section IV-A, the adversarial loss in \eqref{DALoss} satisfies
\begin{equation}
	\begin{aligned}
		\max_{\{D_k\}}(-\mathcal L_{\mathrm{adv}})
		&=-2H(D\mid\widehat Z,\bar Y)\\
		&=-\ln4+2I(D;\widehat Z,\bar Y).
	\end{aligned}
	\label{eq:pseudo_adversarial_information}
\end{equation}
Unlike the ideal case where $I(D;Y)=0$, $I(D;\bar Y)$ may be nonzero because target pseudo-label errors can lead to different label distributions across domains. We next quantify the discrepancy between $I(D;\widehat Z,\bar Y)$ and the ideal class-conditional domain information.

For the full source-target mixture, the pseudo-label error rate is
\begin{equation}
	\delta=\Pr(\bar Y\ne Y).\notag
\end{equation}
Since source labels are exact and the two domains are
equally weighted,
$\delta=\frac12\Pr(\widetilde Y\ne Y\mid D=t)$.
Define
\begin{equation}
	\begin{aligned}
		\xi(\delta)&=h(\delta)+\delta\ln(K-1),\\
		h(\delta)&=-\delta\ln\delta
		-(1-\delta)\ln(1-\delta).
	\end{aligned}
	\label{eq:pseudo_error_function}
\end{equation}

Given $Y$, the label $\bar Y$ has most $K-1$ possible values when $\bar Y\ne Y$.
Hence,
\begin{equation}
	\begin{aligned}
		H(\bar Y\mid Y)\le\xi(\delta),
		H(Y\mid\bar Y)\le\xi(\delta).
	\end{aligned}
	\label{eq:pseudo_entropy_bounds}
\end{equation}
Applying the chain rule of conditional mutual
information and \eqref{eq:pseudo_entropy_bounds}
yields
\begin{equation}
	\left|I(D;\widehat Z\mid Y)
	-I(D;\widehat Z,\bar Y)\right|
	\le \xi(\delta).
	\label{eq:pseudo_cmi_bound}
\end{equation}

This bound quantifies the discrepancy between the ideal CDA information objective and its actual counterpart.

For PCL, let $S_\tau$ denote whether a sample is
retained, where all source samples and only target
samples with confidence $q>\tau$ are selected. The selected pseudo-label error rate and retention rate are
\begin{equation}
	\begin{aligned}
		\delta_\tau=\Pr(\bar Y\ne Y\mid S_\tau=1),
		p_\tau=\Pr(S_\tau=1).\notag
	\end{aligned}
	\label{eq:selected_error_rate}
\end{equation}
Since all source samples are retained, $p_\tau=(1+a_\tau)/2$, where $a_\tau=\Pr(S_\tau=1\mid D=t)$ is the target-domain retention rate. Confidence filtering changes the class proportions.

We use $I_\tau$ and $H_\tau$ to denote information
quantities evaluated under $P_\tau$. The same entropy
argument gives $H_\tau(\bar Y\mid Y)\le\xi(\delta_\tau)$.
Consequently, the class-information term satisfies
\begin{equation}
	\begin{aligned}
		I_\tau(\bar Y;\widehat Z)
		&\le I_\tau(Y,\bar Y;\widehat Z)\\
		&=I_\tau(Y;\widehat Z)
		+I_\tau(\bar Y;\widehat Z\mid Y)\\
		&\le I_\tau(Y;\widehat Z)+\xi(\delta_\tau),
	\end{aligned}
	\label{eq:pseudo_class_information_bound}
\end{equation}

Under independent reference sampling with at least one positive sample per anchor, according to \eqref{eq:pcl_information_lower_bound}, we have
\begin{equation}
	\begin{aligned}
		I_\tau(\bar Y;\widehat Z)
		\ge H_\tau(\bar Y)
		+\mathbb E[\ln|P(i)|]
		-\mathbb E[\mathcal L_{\mathrm{psup}}].
	\end{aligned}
	\label{eq:practical_pcl_information}
\end{equation}
This bound relates the practical PCL loss to the pseudo-label information. Since confidence-based selection changes the distribution used for PCL, we introduce the selection bias
\begin{equation}
	B_\tau=
	\left|I(Y;\widehat Z)
	-I_\tau(Y;\widehat Z)\right|.
	\label{eq:selection_bias}
\end{equation}
By the chain rule, we can get
\begin{equation*}
	\begin{aligned}
		I(Y;\widehat Z)
		=I(Y;\widehat Z\mid S_\tau)
		+I(Y;S_\tau)
		-I(Y;S_\tau\mid\widehat Z).
	\end{aligned}
\end{equation*}
Since $H(S_\tau)=h(p_\tau)$ and
$0\le I(Y;\widehat Z\mid S_\tau=s)\le\ln K$,
we obtain
\begin{equation}
	B_\tau\le h(p_\tau)+(1-p_\tau)\ln K.
	\label{eq:selection_bias_bound}
\end{equation}

Thus, the resulting pseudo-label information objective is
\begin{equation}
	\begin{aligned}
		\mathcal L_{\mathrm{pseudo}}^\tau
		={}R_s+\lambda I(D;\widehat Z,\bar Y)
		-\beta I_\tau(\bar Y;\widehat Z).
	\end{aligned}
	\label{eq:pseudo_information_objective}
\end{equation}
The class-information comparison and the definition of
$B_\tau$ imply
\begin{equation*}
	I(Y;\widehat Z)
	\ge I_\tau(\bar Y;\widehat Z)
	-\xi(\delta_\tau)-B_\tau.
\end{equation*}
Combining this inequality with \eqref{eq:pseudo_cmi_bound}, \eqref{eq:pseudo_class_information_bound}, \eqref{eq:selection_bias}, 
\eqref{eq:selection_bias_bound} and \eqref{eq:pseudo_information_objective}, for $\lambda,\beta\ge0$,
establishes
\begin{equation}
	\begin{aligned}
		\mathcal L_{\mathrm{ideal}}
		\le{}&\mathcal L_{\mathrm{pseudo}}^\tau
		+\lambda\xi(\delta)
		+\beta\xi(\delta_\tau)\\
		&+\beta\left[h(p_\tau)
		+(1-p_\tau)\ln K\right].
	\end{aligned}
	\label{eq:pseudo_objective_bound}
\end{equation}

This bound accounts for both pseudo-label errors
and the selection bias introduced by PCL. The
pseudo-label errors are controlled by $\delta$
for CDA and $\delta_\tau$ for PCL, while the
selection bias depends on the retained sample
proportion. Source pretraining can improve pseudo-label accuracy, while confidence filtering further improves label reliability at the cost of reduced sample coverage, leading to a tradeoff between pseudo-label accuracy and sample coverage.

\subsection{Overall Objective and Training Strategy}
The source classification, class-wise adversarial, and contrastive losses are correspond to the three terms in \eqref{eq:pseudo_information_objective}. The resulting training objective is
\begin{equation}
\mathcal L=\mathcal L_{\mathrm{cls}}
+\lambda\mathcal L_{\mathrm{adv}}
+\beta\mathcal L_{\mathrm{psup}},
\label{TotalLoss}
\end{equation}
 As explained above, the GRL reverses the adversarial gradient reaching the transmitter, while the other losses are minimized normally. The confidence threshold $\tau$ controls the selection of target samples used to form contrastive pairs.

In order to ensure the accuracy of pseudo-labels, we use a two-stage training procedure. First, the network is trained with the labeled source classification loss $\mathcal L_{\mathrm{cls}}$ for 10 epochs to initialize the feature extractor and classifier before target pseudo labels are used. Second, the full objective \eqref{TotalLoss} is optimized with class-wise alignment and contrastive learning. Training uses Adam~\cite{Adam} with learning rate $10^{-4}$ for 60 epochs in total. Target-domain accuracy is used only for evaluation.
\begin{figure}[t]
	\setlength{\belowcaptionskip}{-0.3cm} 
	\vspace{-0.4cm}
	\centering
	\includegraphics[width=3.3in,height=1.2in]{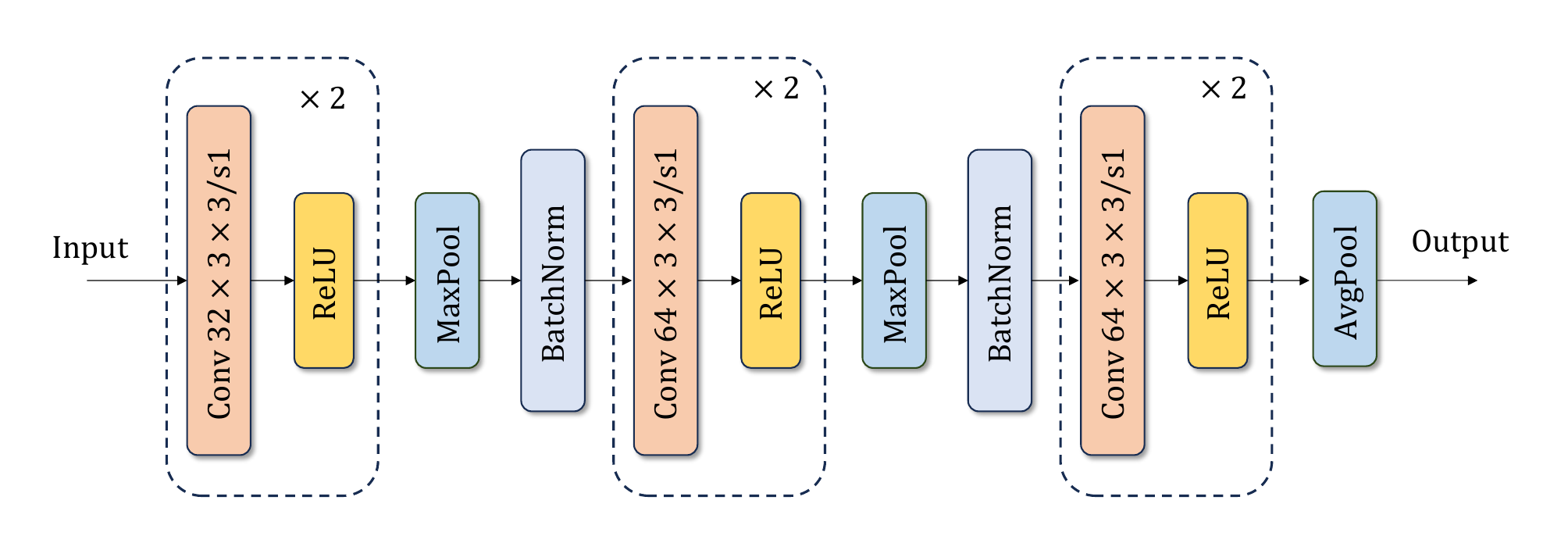}
	\caption{The specific structure of the semantic extractor.}
	\label{FE}
\end{figure}
\begin{table}[t]
	\centering
	\caption{Network architectures of the encoder, decoder, and classifier.}
	\label{encdeccls}
	\renewcommand{\arraystretch}{1.2}
	\begin{tabular}{|c|c|c|c|}
		\hline
		\textbf{Module} & \textbf{Layer Name} & \textbf{Units} & \textbf{Activation} \\
		\hline
		\multirow{5}{*}{\textbf{Encoder}} 
		& Linear & $64$ & ReLU \\\cline{2-4}
		& BatchNorm & -- & -- \\\cline{2-4}
		& Linear & $128$ & ReLU \\\cline{2-4}
		& BatchNorm & -- & -- \\\cline{2-4}
		& Linear & $m$ & -- \\\cline{2-4}
		\hline
		\multirow{5}{*}{\textbf{Decoder}} 
		& Linear & $m$ & ReLU \\\cline{2-4}
		& BatchNorm & -- & -- \\\cline{2-4}
		& Linear & $128$ & ReLU \\\cline{2-4}
		& BatchNorm & -- & -- \\\cline{2-4}
		& Linear & $128$ & -- \\\cline{2-4}
		\hline
		\multirow{3}{*}{\textbf{Classifier}} 
		& Linear & $128$ & ReLU \\\cline{2-4}
		& BatchNorm & -- & -- \\\cline{2-4}
		& Linear & $K$ & Softmax \\\cline{2-4}
		\hline
	\end{tabular}
\end{table}
In this work, the semantic feature extractor adopts specific architectures for different datasets. The value of $\lambda$ in the loss function is set to 0.001, $\beta$ is set to 0.2 and $\tau$ is set to 0.8. For small image datasets, the module is a three-stage convolutional block built from repeated 3$\times$3 kernel and stride 1 convolutions with ReLU  activations, spatial downsampling and batch normalization. The last layer is terminated by a global average pooling. The specific architecture is shown in Fig. \ref{FE}. For large-image datasets, the semantic feature extractor is a pretrained ResNet-18\cite{ResNet}. Besides, the details of the JSCC encoder, decoder, and classifier are shown in Table \ref{encdeccls}. The transmitted dimension is controlled by the number of output units $m$ in the last encoder layer. All the experiments are implemented in PyTorch with a GPU RTX 4060Ti.

\begin{figure*}[t]
	\setlength{\belowcaptionskip}{-0.3cm} 
	\vspace{-0.2cm}
	\subfigcapskip=-5pt
	\centering
	\subfigure[Trained CSNR = 10 dB]{
		\includegraphics[width=2.1in,height=1.4in]{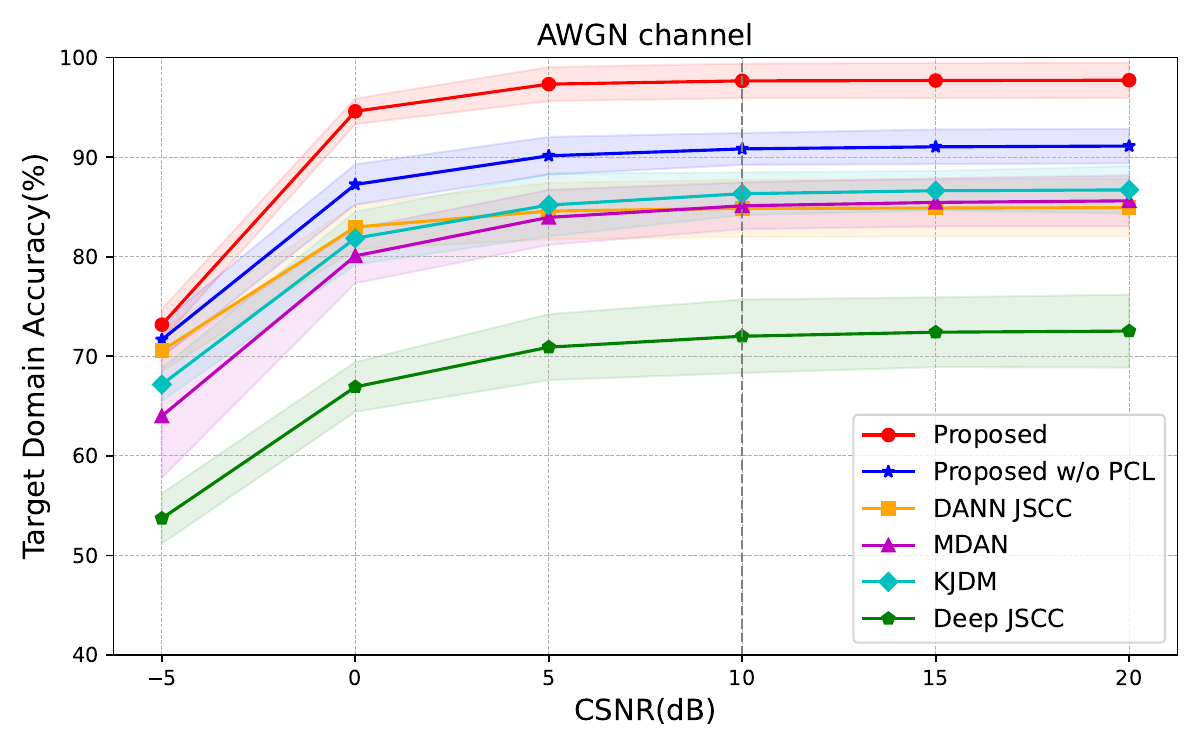}
	}
	\subfigure[Trained CSNR = 15 dB]{
		\includegraphics[width=2.1in,height=1.4in]{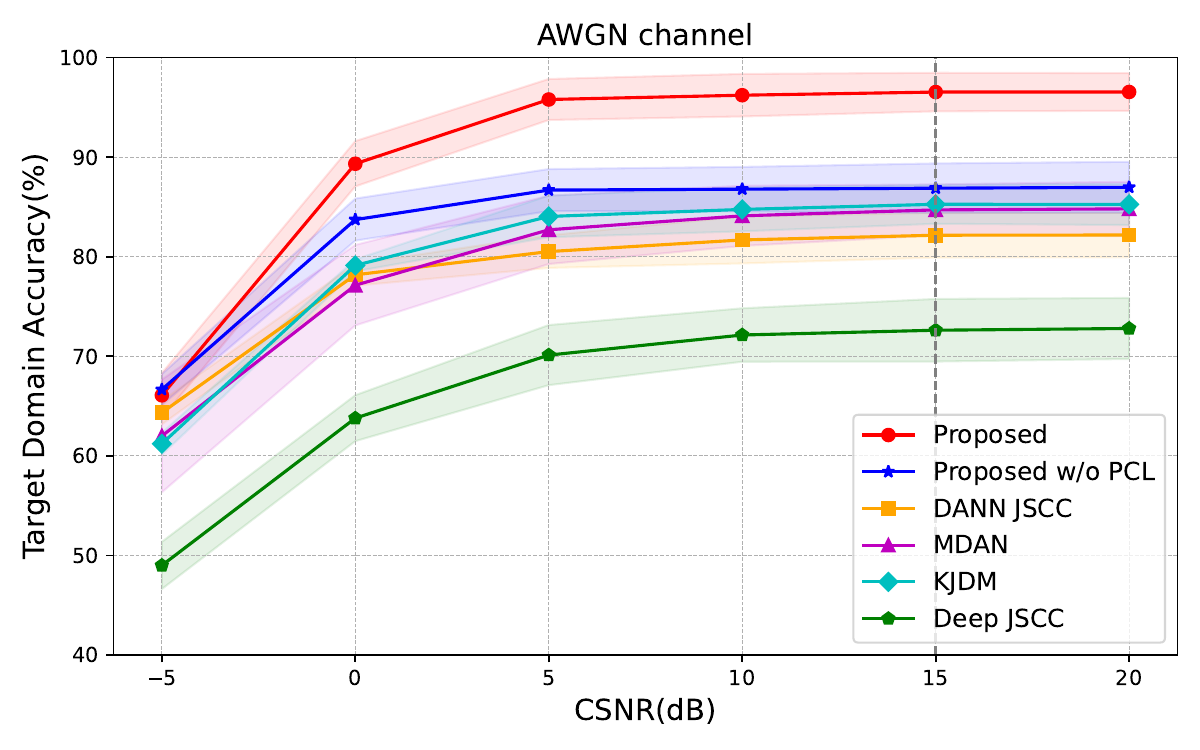}
	}
	\subfigure[Trained CSNR = 10 dB.]{
		\includegraphics[width=2.1in,height=1.4in]{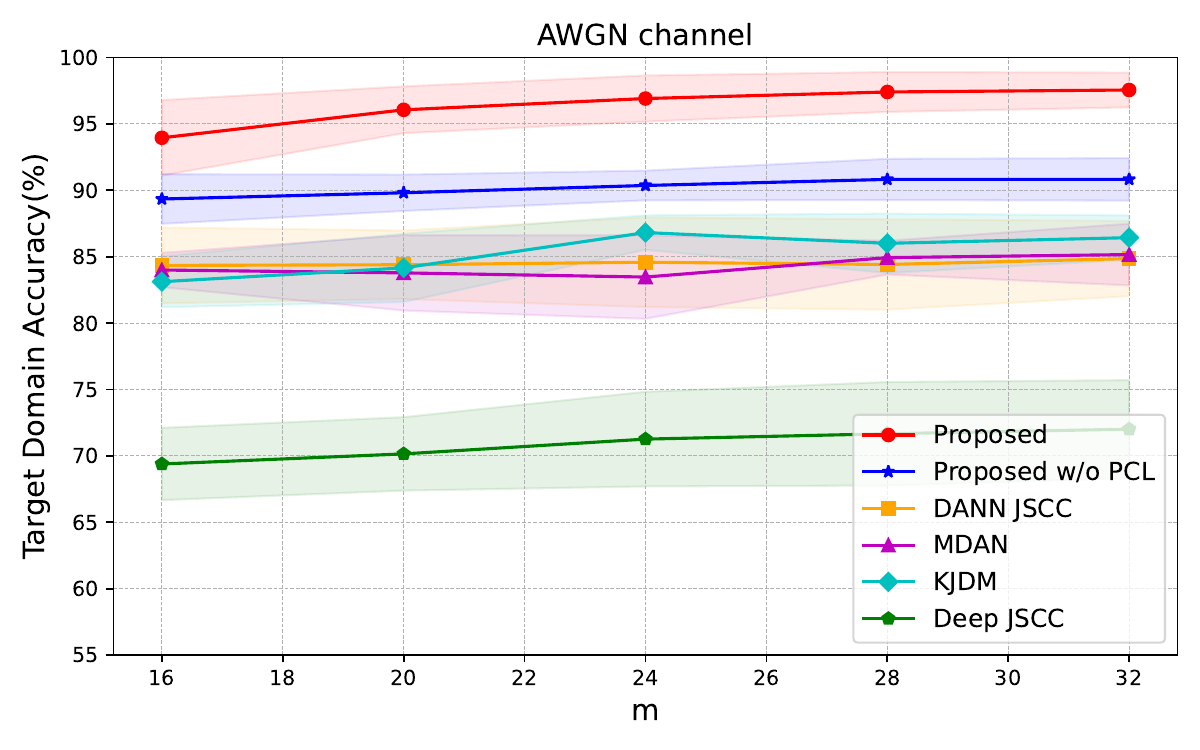}
	}
	\caption{The target-domain classification accuracy for different schemes trained in AWGN channel with fixed CSNR.}
	\label{AWGN}
\end{figure*}

\begin{figure*}[t]
	\setlength{\belowcaptionskip}{-0.3cm} 
	\vspace{-0.2cm}
	\subfigcapskip=-5pt
	\centering
	\subfigure[Trained CSNR = 10 dB]{
		\includegraphics[width=2.1in,height=1.4in]{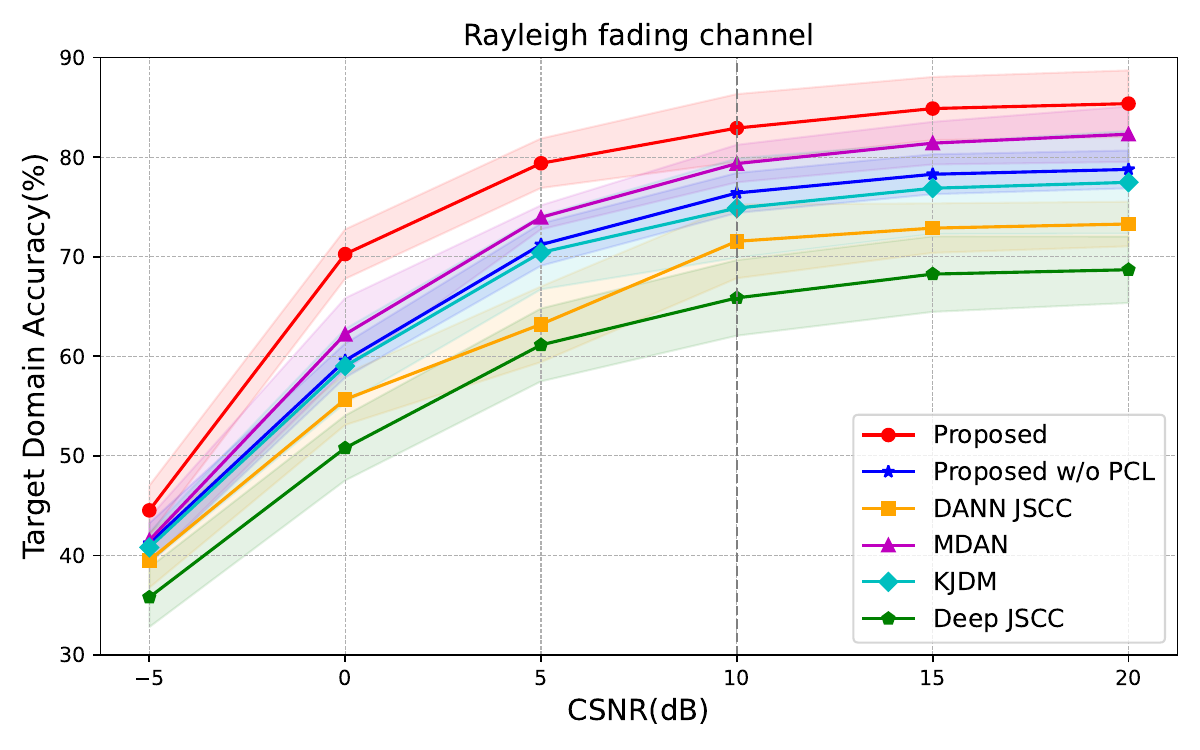}
	}
	\subfigure[Trained CSNR = 15 dB]{
		\includegraphics[width=2.1in,height=1.4in]{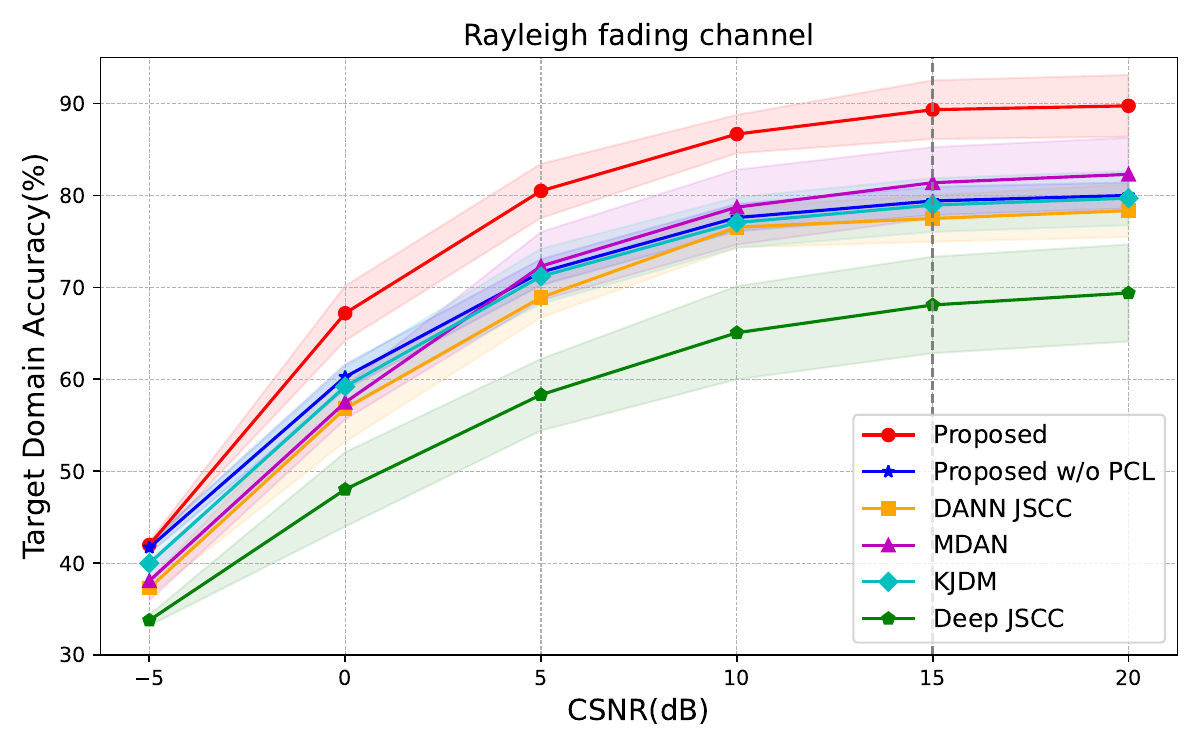}
	}
	\subfigure[Trained CSNR = 10 dB.]{
		\includegraphics[width=2.1in,height=1.4in]{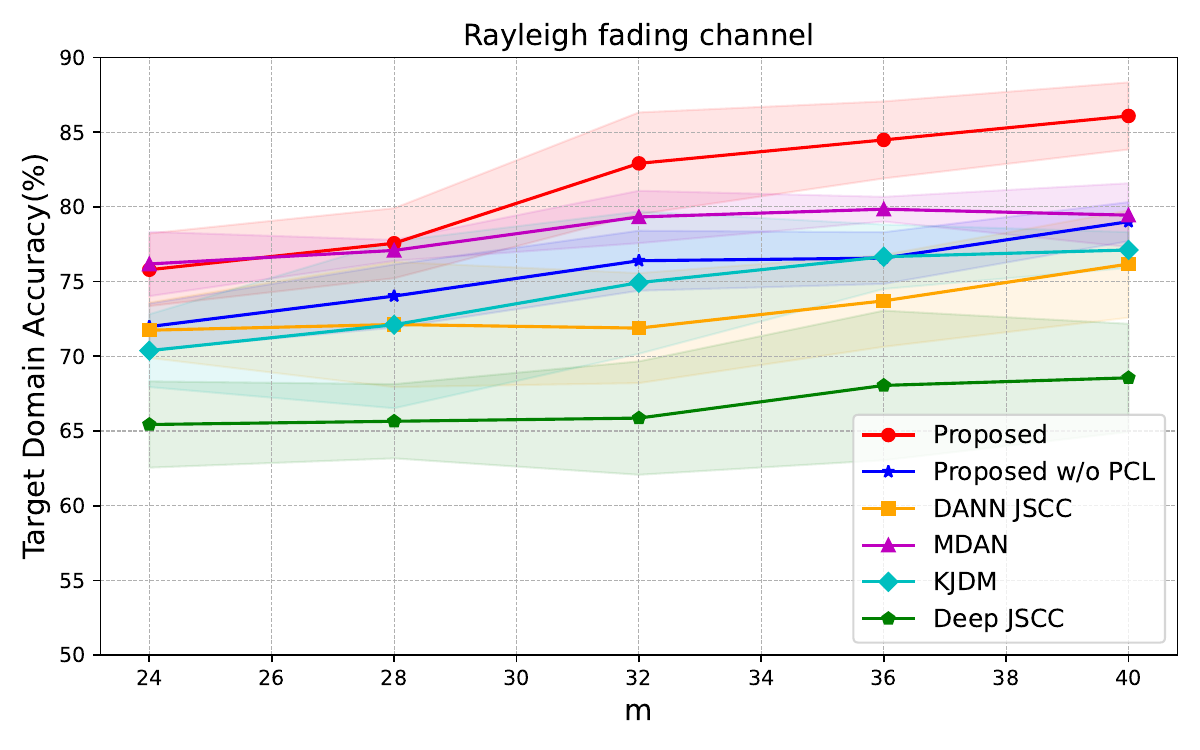}
	}
	\caption{The target-domain classification accuracy for different schemes trained in Rayleigh channel with fixed CSNR.}
	\label{Ray}
\end{figure*}
\section{Experiment}
\subsection{Experimental Setup}
\emph{1) Digits dataset:} We use the Street View House Numbers (SVHN) dataset as the labeled source domain\cite{SVHN}. SVHN  contains 73,257 labeled training images and 26,032 test images. All images are 32$\times$32 RGB. For the unlabeled target domain, we adopt MNIST dataset\cite{MNIST}. MNIST consists of grayscale images of handwritten digits, with 28$\times$28 resolution. Note that MNIST test labels are reserved for evaluation only. For compatibility, MNIST images are resized to 32$\times$32 and replicated across three channels to match RGB inputs. Relative to SVHN, MNIST has cleaner backgrounds, centered digits, and lower visual complexity, which leads to a pronounced domain shift in color, texture and background. This induces a significant appearance gap between source and target, which is commonly used for assessing cross-domain generalization in task-oriented Deep JSCC. 

\emph{2) PACS dataset:} PACS is a multi-domain visual dataset consisting of four stylistic domains: photo (P), art painting (A), cartoon (C) and sketch (S)\cite{PACS}. Each domain has seven categories (dog, elephant, giraffe, guitar, horse, house and person). With 8,977 training images and 1,014 test images, PACS is larger and  more complex than digit datasets. All images are resized to a common resolution 224$\times$224 by central cropping. Because domain adaptation is considered, we need to select one domain as the labeled source domain and a different domain as the unlabeled target domain.  Only the target’s images without labels are accessible during training and target labels are used for evaluation only.

\begin{figure*}[t]
	\setlength{\belowcaptionskip}{-0.3cm} 
	\vspace{-0.2cm}
	\subfigcapskip=-5pt
	\centering
	\subfigure[Deep JSCC]{
		\includegraphics[width=1.9in,height=1.8in]{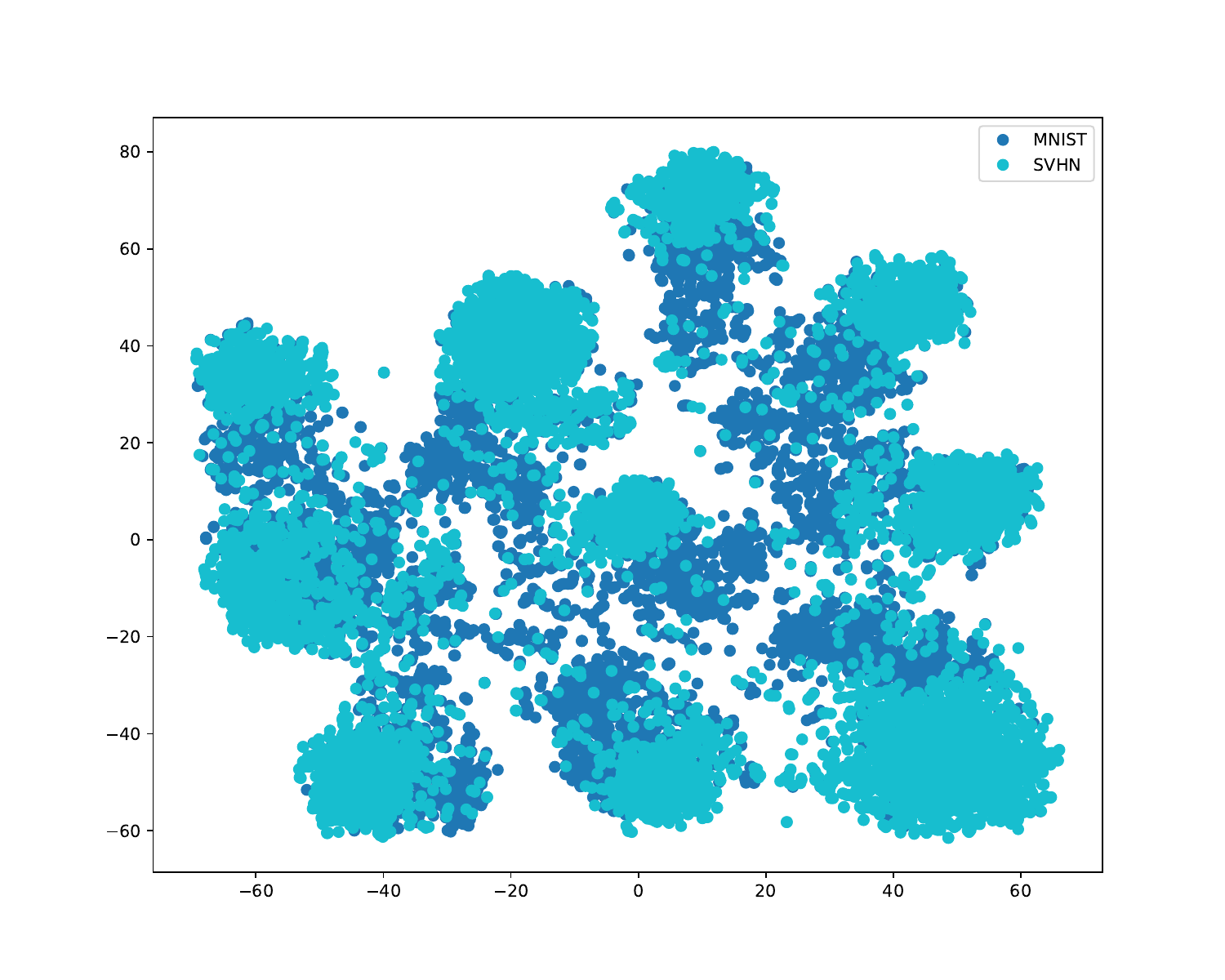}
	}
	\subfigure[DANN JSCC]{
		\includegraphics[width=1.9in,height=1.8in]{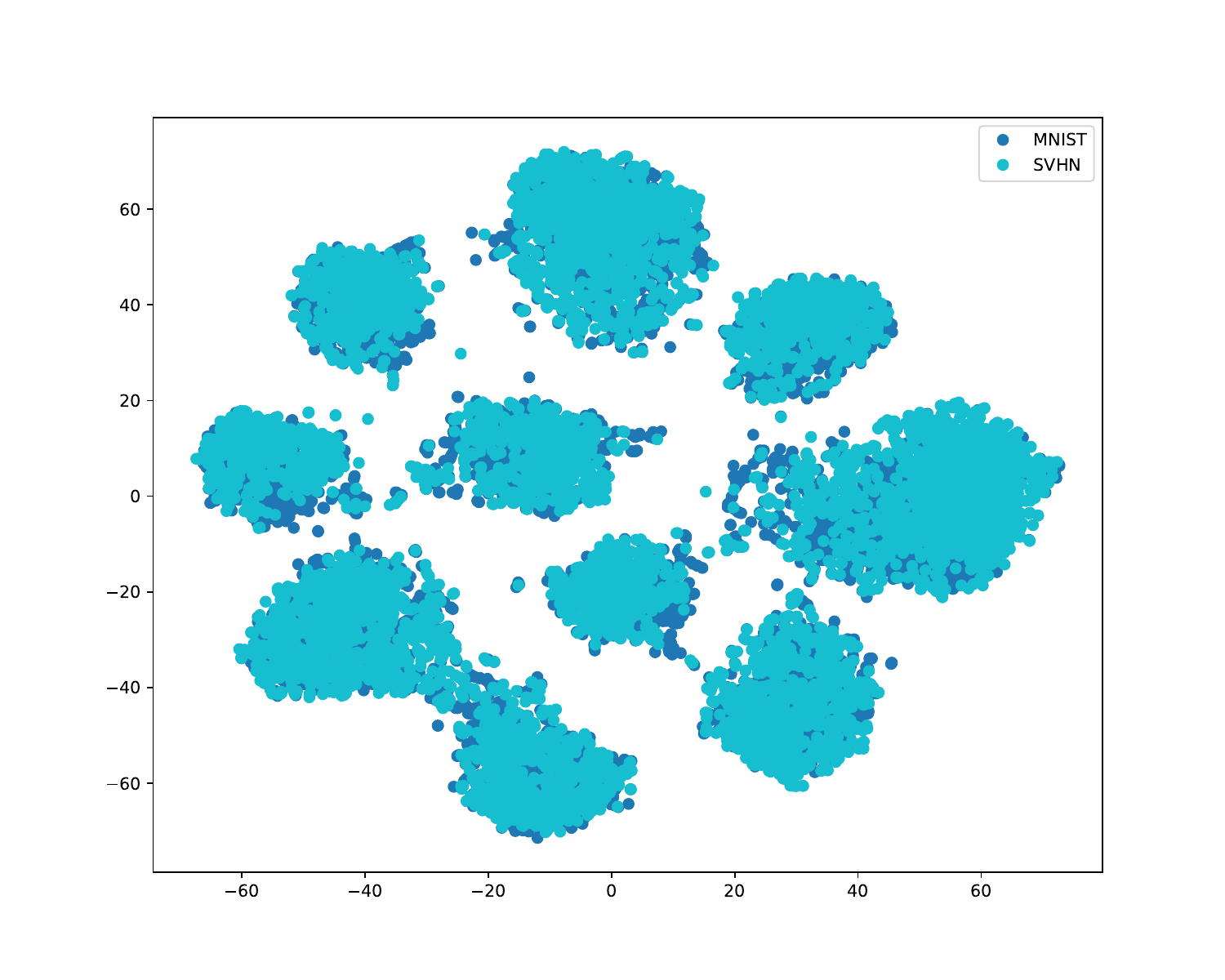}
	}
	\subfigure[Proposed]{
		\includegraphics[width=1.9in,height=1.85in]{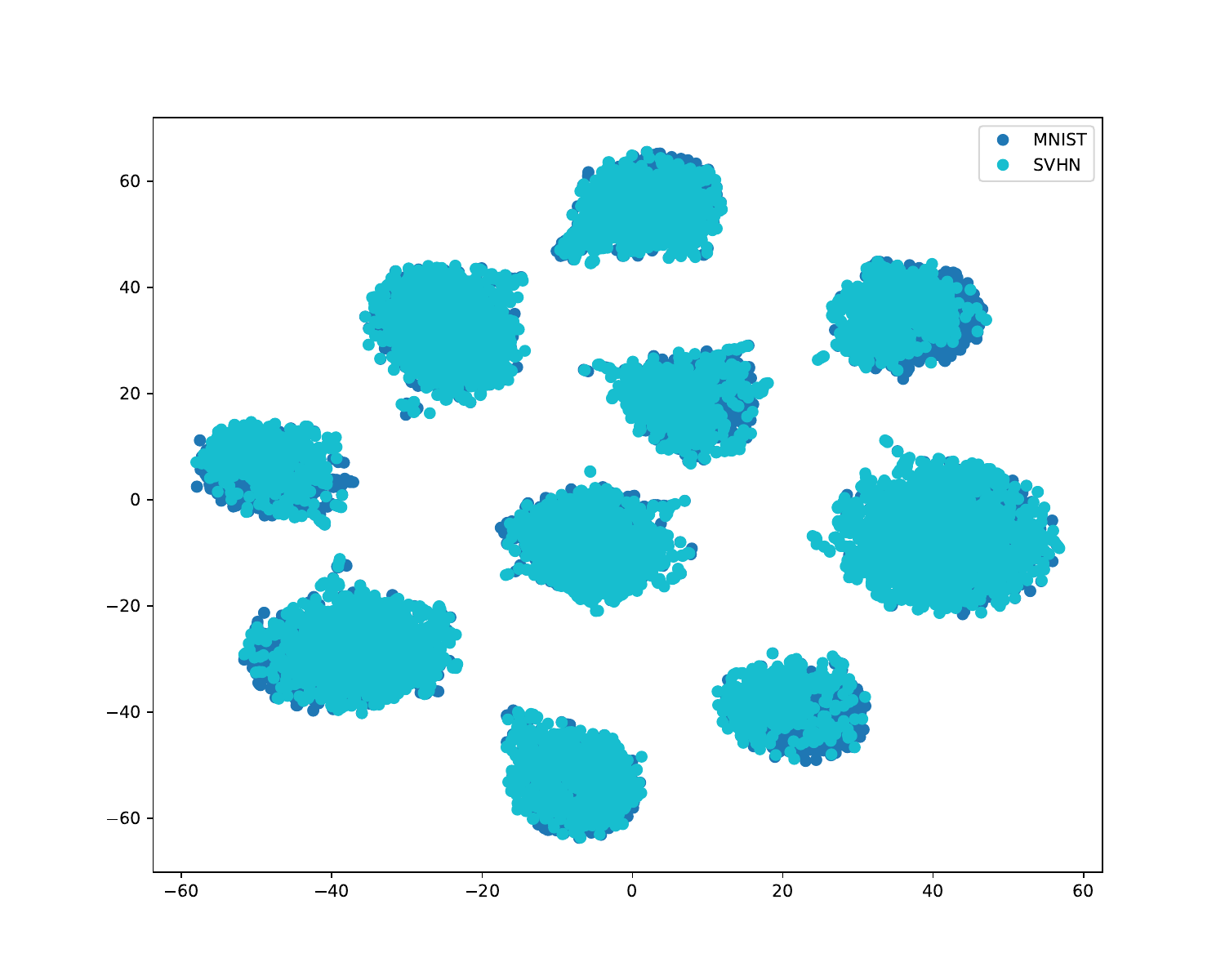}
	}
	\caption{The t-SNE visualization of different schemes.}
	\label{tSNE}
\end{figure*}

\emph{3) Comparison schemes.} 
We compare the proposed method with the conventional Deep JSCC scheme~\cite{djscc1} and the state-of-the-art domain-adaptive JSCC method MDAN~\cite{OOD4}. The conventional Deep JSCC scheme does not perform any domain adaptation. It is trained only on the source domain and directly evaluated on the target domain. MDAN transfers target-domain samples to the source-domain style through StarGAN, and therefore requires an additional StarGAN module during inference. We also compare our method with DANN-based domain-adaptive JSCC~\cite{DANN}, which performs global feature-level domain alignment through adversarial learning. In addition, KJDM \cite{KMUR} is adopted as a recent domain adaptation baseline. Unlike global marginal feature alignment, KJDM matches the source feature-label joint distribution with the pseudo-labeled target joint distribution, thereby introducing class-aware distribution alignment.

\begin{figure*}[!t]
	\setlength{\belowcaptionskip}{-0.3cm} 
	\vspace{-0.2cm}
	\centering
	\includegraphics[width=6.8in,height=1.7in]{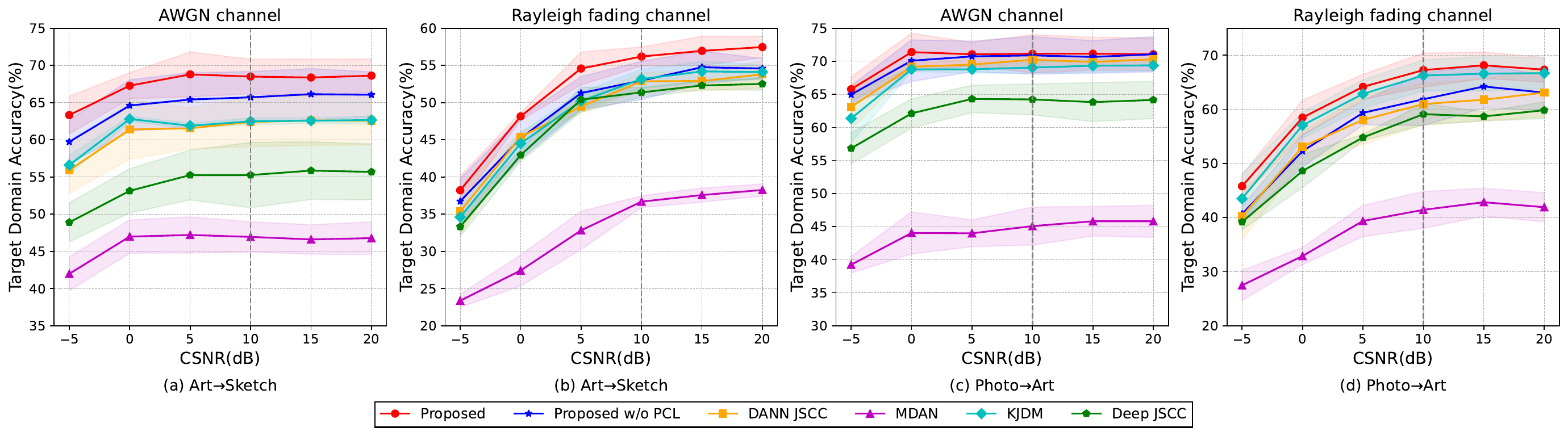}
	\caption{Target-domain classification accuracy versus test CSNR for Art-to-Sketch and Photo-to-Art adaptation under AWGN and Rayleigh fading channels. The models are trained at a CSNR of 10 dB.}
	\label{pacs_AS_PA}
\end{figure*}
\begin{figure*}[!t]
	\setlength{\belowcaptionskip}{-0.3cm} 
	\vspace{-0.2cm}
	\centering
	\includegraphics[width=6.8in,height=1.7in]{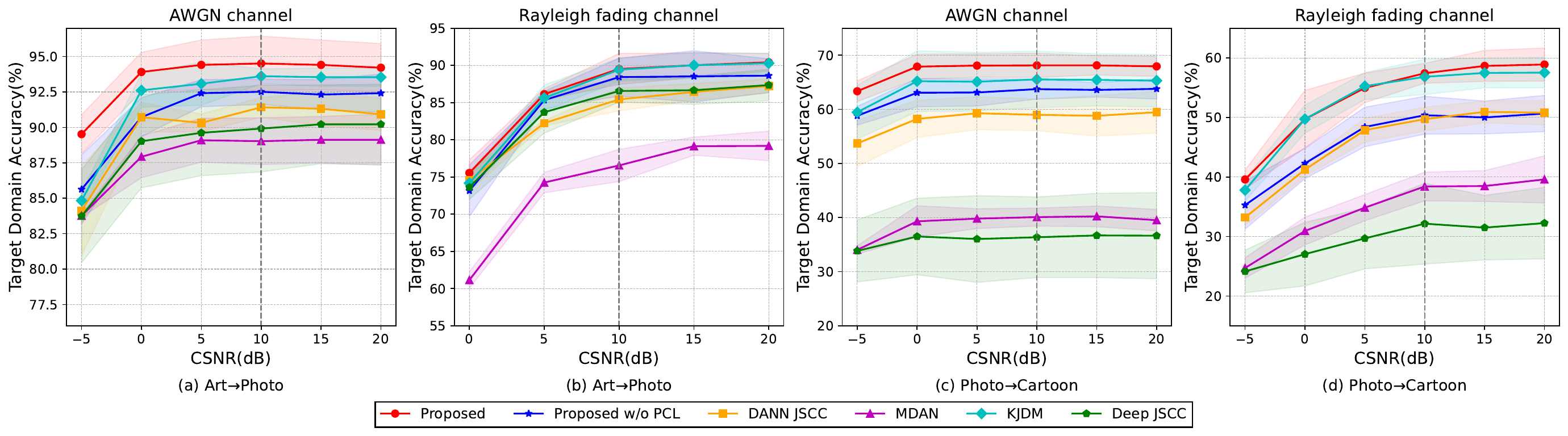}
	\caption{Target-domain classification accuracy versus test CSNR for Art-to-Photo and Photo-to-Cartoon adaptation under AWGN and Rayleigh fading channels. The models are trained at a CSNR of 10 dB.}
	\label{pacs_AP_PC}
\end{figure*}
\begin{figure*}[!t]
	\setlength{\belowcaptionskip}{-0.3cm} 
	\vspace{-0.2cm}
	\centering
	\includegraphics[width=6.8in,height=1.7in]{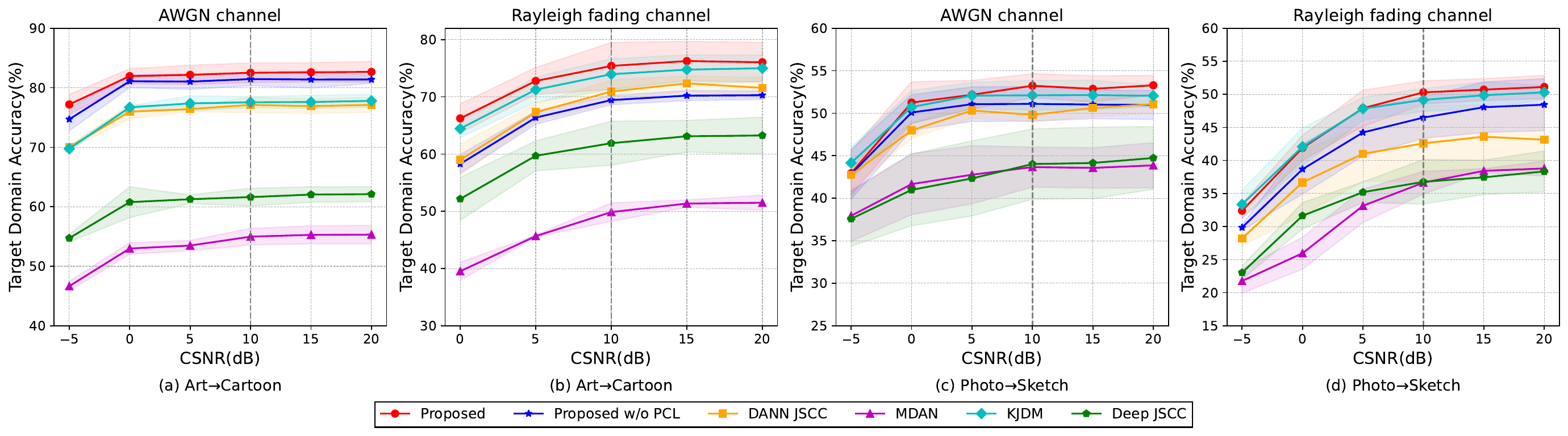}
	\caption{Target-domain classification accuracy versus test CSNR for Art-to-Cartoon and Photo-to-Sketch adaptation under AWGN and Rayleigh fading channels. The models are trained at a CSNR of 10 dB.}
	\label{pacs_AC_PS}
\end{figure*}
\subsection{Experimental Results on Digits}
We first conduct the experiments on the digits datasets. We utilize target-domain classification accuracy to assess and compare the different methods. According to Section III, CSNR constrains the label information available in the received signal, while received-signal adaptation controls conditional domain information. We vary test CSNR to evaluate channel robustness. We also vary the transmitted dimension $m$ as an empirical communication-resource sensitivity study; because changing $m$ changes the learned representation and requires retraining, this sweep is not presented as validation of a theoretical monotonic law in $m$. All experiments are repeated with multiple random seeds. The curves show the mean performance, and the shaded bands represent $\pm 1$ standard deviation across seeds.

In Fig.~\ref{AWGN}, we present the results of different methods across various test CSNRs under the AWGN channel. The models are trained at fixed CSNRs of 10 and 15 dB and evaluated over a wider CSNR range, while the channel-output dimension is fixed at $m=32$. The Deep JSCC, which does not employ any domain adaptation operation, generally exhibits the lowest target-domain accuracy because it cannot mitigate the distribution discrepancy between the source and target domains. DANN JSCC improves the target-domain generalization performance through global feature alignment, but its global discriminator cannot explicitly capture class-dependent domain discrepancies. KJDM further reduces the domain gap by matching the joint distributions of the source and target representations and generally performs better than vanilla Deep JSCC and DANN JSCC in several settings. Nevertheless, it remains inferior to the proposed method. MDAN also improves upon the conventional baselines, but its performance remains substantially below that of the proposed method. Benefiting from class-level feature alignment and pseudo-label supervised contrastive learning, the proposed method achieves the best overall results. At the corresponding training CSNRs of 10 and 15~dB, it attains target-domain classification accuracies of 98.15\% and 96.71\%, respectively. To further identify the contribution of pseudo-label supervised contrastive learning, we also compare the proposed method with its variant without PCL.

Besides, we further evaluate all methods under a Rayleigh fading channel, as shown in Fig. \ref{Ray}. The proposed method still substantially improves the target-domain classification accuracy and consistently outperforms all compared schemes. When CSNR is 10 dB and 15 dB, it attains target-domain accuracies of 82.34\% and 89.98\%, respectively. In this setting, MDAN outperforms the variant of our method without pseudo-label–based contrastive learning. However, MDAN requires an additional generative network at inference, incurring additional computational and space overhead. 

Fig. \ref{AWGN} and Fig. \ref{Ray} also illustrate the target-domain classification accuracy versus the channel output dimension m for different methods, trained at CSNR = 10 dB. The performance of MDAN is still inferior to the proposed scheme, while DANN JSCC and Deep JSCC lag behind.

Fig. \ref{tSNE} shows the t-SNE visualization of the channel outputs of different schemes.  The proposed method yields well-separated and compact class clusters in which source and target received signals are closely overlapped, which is qualitatively consistent with greater domain invariance and class discriminability in received representations.

\begin{table*}[t]
	\centering
	\caption{Component and adaptation-location ablation on Digits at training and test CSNRs of 10 dB.}
	\label{tab:ablation_10db}
	\renewcommand{\arraystretch}{1.15}
	\begin{tabular*}{\textwidth}
		{@{\extracolsep{\fill}}lcccccc@{}}
		\toprule
		Method & CDA & PCL & Position
		& Source Acc.(\%) & Target Acc.(\%) & Target gain (\%) \\
		\midrule
		Source-only (Deep JSCC)
		& -- & -- & --
		& $93.42 \pm 0.24$ & $71.38 \pm 3.59$ & $0.00$ \\
		Tx-CDA
		& $\checkmark$ & -- & Tx
		& $93.73 \pm 0.08$ & $86.34 \pm 2.08$ & $+14.96$ \\
		Rx-CDA
		& $\checkmark$ & -- & Rx
		& $93.62 \pm 0.21$ & $91.05 \pm 2.26$ & $+19.67$ \\
		Tx-PCL
		& -- & $\checkmark$ & Tx
		& $93.99 \pm 0.38$ & $95.24 \pm 0.53$ & $+23.86$ \\
		Rx-PCL
		& -- & $\checkmark$ & Rx
		& $94.04 \pm 0.20$ & $96.89 \pm 0.45$
		& $+25.51$ \\
		Tx-CDA+PCL
		& $\checkmark$ & $\checkmark$ & Tx
		& $94.10 \pm 0.33$ & $95.44 \pm 1.06$ & $+24.06$ \\
		Rx-CDA+PCL (Proposed)
		& $\checkmark$ & $\checkmark$ & Rx
		& $93.96 \pm 0.38$ & $\mathbf{98.15 \pm 1.23}$ & $\mathbf{+26.77}$ \\
		\bottomrule
	\end{tabular*}
\end{table*}

\begin{table*}[t]
	\centering
	\caption{Component ablation on PACS with Art as the
		source domain at training and test CSNRs of 10 dB.}
	\label{tab:pacs_art_ablation}
	\small
	\renewcommand{\arraystretch}{1.15}
	\setlength{\tabcolsep}{3pt}
	\begin{tabular*}{\textwidth}
		{@{\extracolsep{\fill}}lcccccccc@{}}
		\toprule
		& \multicolumn{3}{c}{AWGN}
		& \multicolumn{3}{c}{Rayleigh} \\
		\cmidrule(lr){2-4}
		\cmidrule(lr){5-7}
		Method
		& A$\rightarrow$C & A$\rightarrow$P
		& A$\rightarrow$S 
		& A$\rightarrow$C & A$\rightarrow$P
		& A$\rightarrow$S  \\
		\midrule
		Source-only (Deep JSCC)
		& $63.66 \pm 1.87$
		& $89.92 \pm 2.06$
		& $56.31 \pm 3.96$
		& $62.29 \pm 2.24$
		& $87.23 \pm 1.54$
		& $53.09 \pm 2.65$ \\ 
		CDA-only
		& $81.91 \pm 2.68$
		& $92.51 \pm 0.90$
		& $66.46 \pm 3.44$
		& $70.44 \pm 1.69$
		& $88.42 \pm 2.55$
		& $52.91 \pm 2.46$ \\
		PCL-only
		& $82.32 \pm 1.67$
		& $93.01 \pm 1.05$
		& $48.01 \pm 3.02$
		& $67.24 \pm 3.76$
		& $87.72 \pm 3.53$
		& $42.34 \pm 4.52$ \\
		CDA+PCL (Proposed)
		& $\mathbf{82.40 \pm 0.56}$
		& $\mathbf{94.51 \pm 3.92}$
		& $\mathbf{68.64 \pm 2.81}$
		& $\mathbf{73.78 \pm 3.40}$
		& $\mathbf{89.52 \pm 2.07}$
		& $\mathbf{56.40 \pm 2.39}$ \\
		\bottomrule
	\end{tabular*}
\end{table*}

\subsection{Experimental Results on PACS}
Figs. \ref{pacs_AS_PA}--\ref{pacs_AC_PS} report the target-domain classification accuracy on the PACS dataset under AWGN and Rayleigh fading channels. We consider six representative domain adaptation tasks. Both Art and Photo are considered as source domains, while the remaining three domains are separately used as target domains. The dimension of the channel output is set to $m=128$. All models are trained at a CSNR of 10 dB and evaluated over a wide range of test CSNRs.

As shown in Fig. \ref{pacs_AS_PA}, Fig. \ref{pacs_AP_PC} and Fig. \ref{pacs_AC_PS}, the proposed method generally achieves the highest target-domain classification accuracy across all six adaptation tasks. Under the AWGN channel and at the training CSNR of 10 dB, the proposed method achieves approximately 68.64\%, 72.12\%, 94.51\%, 68.13\%, 82.40\%, and 53.23\% on A$\rightarrow$S, P$\rightarrow$A, A$\rightarrow$P, P$\rightarrow$C, A$\rightarrow$C, and P$\rightarrow$S, respectively. The results also reveal substantial differences in adaptation difficulty among different domain pairs. For example, A$\rightarrow$P achieves relatively high accuracy because the semantic and visual discrepancy between the two domains is comparatively limited, whereas P$\rightarrow$S remains more challenging due to the pronounced appearance difference between natural photographs and sketch images. Nevertheless, the proposed method maintains a clear advantage over the compared schemes in both relatively easy and difficult adaptation settings. Compared with the AWGN results, Rayleigh fading generally leads to lower classification accuracy and larger performance degradation at low CSNRs because both additive noise and random channel fading affect the transmitted representations. Despite these more challenging channel conditions, the proposed method continues to outperform Deep JSCC, DANN JSCC, MDAN, and KJDM across the considered domain pairs, demonstrating its robustness to both distribution shifts and channel variations.

Moreover, the proposed method generally outperforms its variant without pseudo-label supervised contrastive learning. This advantage indicates that PCL provides useful additional training constraints beyond pseudo-label-based class-wise alignment in the considered finite-sample settings. By encouraging same-label cross-domain samples to be similar and differently labeled samples to be separated, PCL can improve the discriminative geometry of the received representations. Overall, the consistent gains over all source-target pairs show that the proposed framework is not restricted to a particular domain combination and can generalize to diverse cross-domain transmission scenarios.

\begin{figure}[t]
	\setlength{\belowcaptionskip}{-0.3cm} 
	\vspace{-0.2cm}
	\subfigcapskip=-5pt
	\centering
	\subfigure[Deep JSCC]{
		\includegraphics[width=1.5in,height=1.4in]{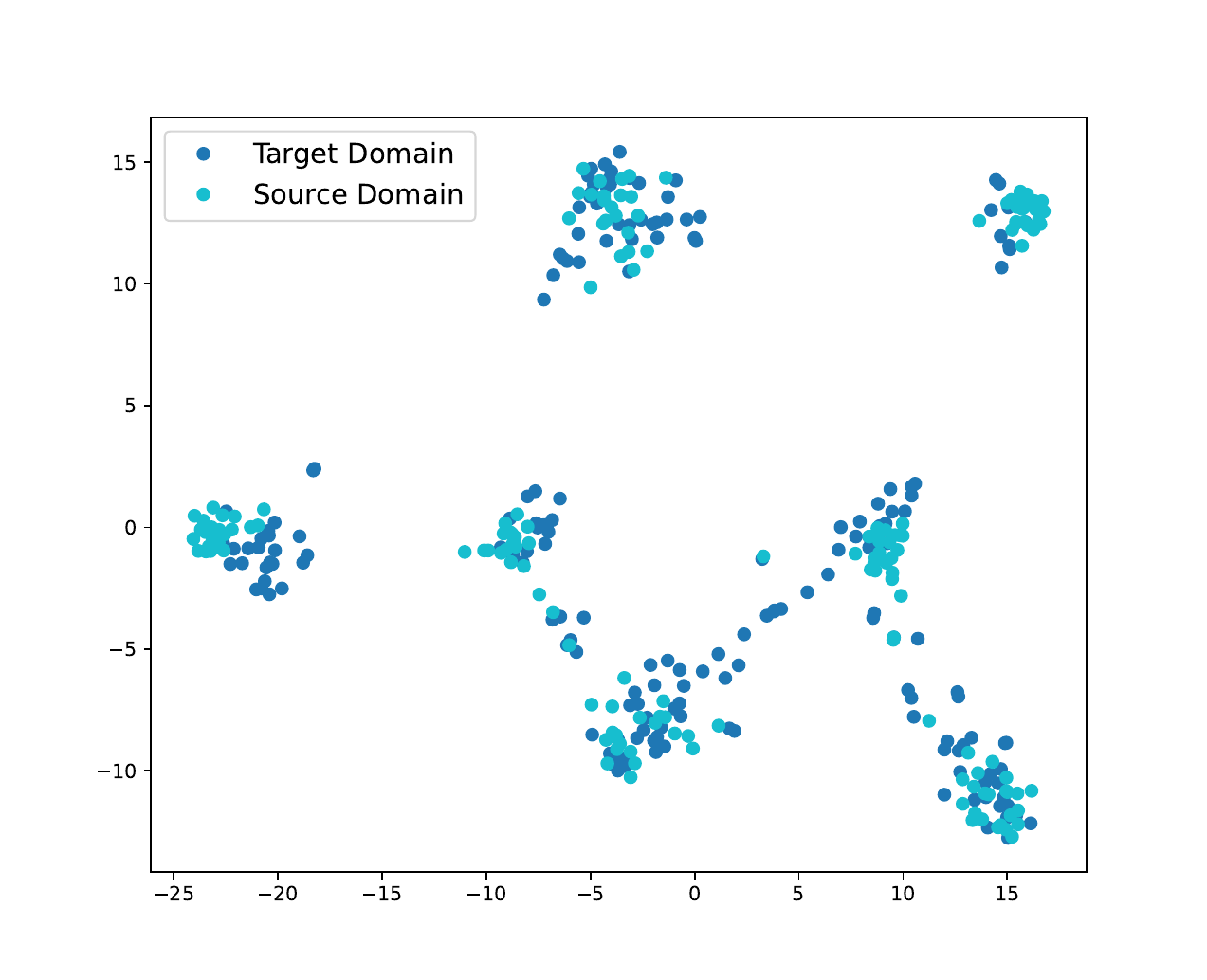}
	}
	\subfigure[Proposed]{
		\includegraphics[width=1.5in,height=1.4in]{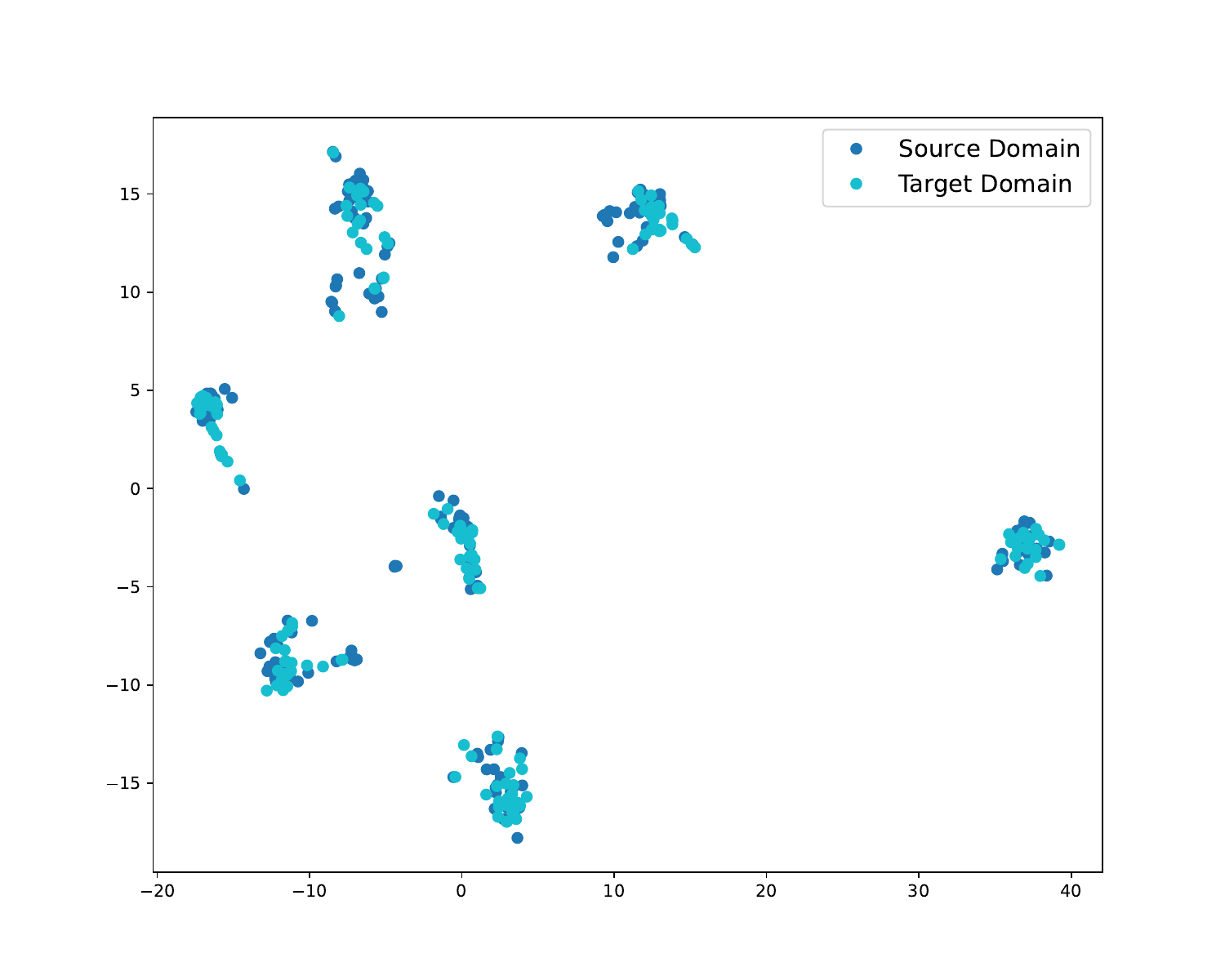}
	}
	\caption{The t-SNE visualization of different schemes on PACS dataset.}
	\label{AC_tNSE}
\end{figure}

Besides, the t-SNE visualization of the channel outputs under the PACS dataset is conducted. As shown in Fig. \ref{AC_tNSE}, our proposed method yields greater overlap between same-class samples across different domains, forms more compact class clusters, and produces more pronounced separation between different class clusters.
\subsection{Ablation Study}
Table~\ref{tab:ablation_10db} examines the contributions
of CDA and PCL and their adaptation locations.
Both modules improve target accuracy over Source-only,
and receiver-side (RX) adaptation outperforms its
transmitter-side (TX) counterpart in all three configurations.
At the receiver, CDA and PCL individually achieve
91.05\% and 96.89\% target accuracy, respectively.
Combining them increases accuracy to 98.15\%.
These results support applying both adaptation losses
to the received representations used for classification.
Receiver-side adaptation achieves higher target accuracy
and directly constrains the channel-impaired representations
used for classification. Table~\ref{tab:pacs_art_ablation} further reports an ablation study of CDA and PCL on PACS, where their combination also achieves higher accuracy than either module alone, supporting their complementary roles.

\begin{table*}[t]
	\centering
	\caption{Target-domain accuracy under different parameter settings on A$\rightarrow$P under AWGN channel, with training and test CSNRs of 10 dB.}
	\label{tab:ap_hyperparameter}
	\renewcommand{\arraystretch}{1.2}
	\begin{tabular*}{\textwidth}
		{@{\extracolsep{\fill}}cccccc@{}}
		\toprule
		\multicolumn{2}{c}{$\lambda$ sweep}
		& \multicolumn{2}{c}{$\beta$ sweep}
		& \multicolumn{2}{c}{$\tau$ sweep} \\
		\multicolumn{2}{c}{$\beta=0.2,\ \tau=0.8$}
		& \multicolumn{2}{c}{$\lambda=0.001,\ \tau=0.8$}
		& \multicolumn{2}{c}{$\lambda=0.001,\ \beta=0.2$} \\
		\cmidrule(lr){1-2}
		\cmidrule(lr){3-4}
		\cmidrule(lr){5-6}
		$\lambda$ & Target Acc. (\%)
		& $\beta$ & Target Acc. (\%)
		& $\tau$ & Target Acc. (\%) \\
		\midrule
		$0$       & $93.01 \pm 1.05$
		& $0$     & $92.51 \pm 0.90$
		& $0.50$  & $94.01 \pm 1.87$ \\
		
		$10^{-4}$ & $92.36 \pm 1.20$
		& $0.10$  & $94.05 \pm 1.83$
		& $0.60$  & $94.21 \pm 1.35$ \\
		
		$10^{-3}$ & $\mathbf{94.51 \pm 3.92}$
		& $0.20$  & $\mathbf{94.51 \pm 3.92}$
		& $0.70$  & $\mathbf{94.51 \pm 1.75}$ \\
		
		$10^{-2}$ & $91.42 \pm 1.35$
		& $0.50$  & $91.62 \pm 3.96$
		& $0.80$  & $\mathbf{94.51 \pm 3.92}$ \\
		
		$10^{-1}$ & $93.01 \pm 1.95$
		& $1.00$  & $91.12 \pm 2.42$
		& $0.90$  & $93.01 \pm 3.23$ \\
		\bottomrule
	\end{tabular*}
\end{table*}
\begin{table}[t]
	\centering
	\caption{Domain discrepancy and target class separability
		 on Digits under AWGN channel, with a test CSNR of 10 dB. All metrics are computed
		on received representations.}
	\label{tab:mmd_fisher}
	\renewcommand{\arraystretch}{1.2}
	\setlength{\tabcolsep}{4pt}
	\begin{tabular}{lccc}
		\toprule
		Method
		& Conditional MMD$^2$ $\downarrow$
		& Target Fisher $\uparrow$ \\
		\midrule
		Source-only
		& $0.1090 \pm 0.0061$
		& $0.838 \pm 0.102$ \\
		CDA
		& $0.0417 \pm 0.0046$
		& $1.340 \pm 0.069$ \\
		CDA+PCL (proposed)
		&  $\mathbf{0.0129 \pm 0.0025}$
		& $\mathbf{3.510 \pm 0.389}$ \\
		\bottomrule
	\end{tabular}
\end{table}

We evaluate the sensitivity to parameters $\lambda$, $\beta$, and $\tau$, as reported in Table~\ref{tab:ap_hyperparameter}. With $\beta=0.2$ and $\tau=0.8$, the highest mean target accuracy of $94.51\%$ is achieved at $\lambda=0.001$. Larger alignment weights reduce accuracy, indicating that stronger emphasis on alignment does not necessarily improve target-domain classification. With $\lambda=0.001$ and $\tau=0.8$, increasing $\beta$ from zero to $0.2$ improves the mean target accuracy from $92.51\%$ to $94.51\%$. Further increasing $\beta$ reduces accuracy, supporting a moderate contribution from contrastive learning and conditional alignment. For confidence filtering, $\tau=0.7$ and $\tau=0.8$ achieve the same highest mean accuracy. Lower thresholds admit more target samples but may introduce incorrect pseudo labels, whereas a high threshold can limit the samples available for contrastive learning. The results suggest balancing pseudo-label reliability and sample coverage, which corresponds to the analysis in \eqref{eq:pseudo_objective_bound}.

We then evaluate conditional domain discrepancy using
class-conditional MMD$^2$,
\begin{equation}
	\Delta_{\mathrm{MMD}}
	=
	\frac{1}{K}\sum_{k=1}^{K}
	\operatorname{MMD}^2(P_{s,k},P_{t,k}),
\end{equation}
where $P_{s,k}$ and $P_{t,k}$ are the received-feature
distributions of class $k$ in the two domains.
A smaller value indicates closer class-conditional
distributions across domains. Target class separability is measured by the Fisher ratio,
\begin{equation}
	F=
	\frac{\sum_{k=1}^{K}n_k\|\mu_k-\mu\|^2}
	{\sum_{k=1}^{K}\sum_{i:y_i=k}
		\|\widehat z_i-\mu_k\|^2},
\end{equation}
where $n_k$ and $\mu_k$ are the target sample number and
received-feature mean of class $k$, and $\mu$ is the
overall target mean. A larger ratio indicates greater
between-class separation relative to within-class spread.
These metrics assess domain alignment and class discriminability.

\begin{table}[t]
	\centering
	\renewcommand{\arraystretch}{1.3}
	\caption{High-confidence pseudo-label accuracy and coverage under the severe A$\rightarrow$S domain shift on PACS.}
	\label{tab:pseudo_label_quality}
	\begin{tabular}{llcc}
		\hline
		Channel & Training strategy
		& Pseudo-label Acc.(\%) & Cover. (\%) \\
		\hline
		\multirow{2}{*}{AWGN}
		& Single-stage & 65.29 & 87.36 \\
		& Two-stage    & \textbf{68.67} & \textbf{90.22} \\
		\hline
		\multirow{2}{*}{Rayleigh}
		& Single-stage & 54.82 & 63.38 \\
		& Two-stage    & \textbf{58.97} & \textbf{67.33} \\
		\hline
	\end{tabular}
\end{table}

\begin{table*}[t]
	\centering
	\caption{Effect of confidence filtering on target-domain
		accuracy (\%) on PACS under an AWGN channel.}
	\label{tab:pacs_confidence_filter_ablation}
	\renewcommand{\arraystretch}{1.15}
	\begin{tabular*}{\textwidth}{@{\extracolsep{\fill}}lcccccc@{}}
		\toprule
		\multirow{2}{*}{Method}
		& \multicolumn{2}{c}{Filtering}
		& \multicolumn{4}{c}{Target Acc. (\%)} \\
		\cmidrule(lr){2-3}\cmidrule(lr){4-7}
		& CDA & PCL & A$\to$C & A$\to$P & A$\to$S & Avg. \\
		\midrule
		\multirow{4}{*}{CDA + PCL}
		& $\times$ & $\times$
		& $81.45 \pm 2.72$
		& $93.61 \pm 1.95$
		& $67.56 \pm 4.48$
		& $80.87$ \\
		& $\checkmark$ & $\times$
		& $75.91 \pm 2.37$
		& $91.92 \pm 0.30$
		& $66.00 \pm 3.74$
		& $77.94$ \\
		& $\times^{*}$ & $\checkmark^{*}$
		& $\mathbf{82.40 \pm 0.56}$
		& $94.51 \pm 3.92$
		& $\mathbf{68.64 \pm 2.81}$
		& $\mathbf{81.85}$ \\
		& $\checkmark$ & $\checkmark$
		& $82.02 \pm 0.98$
		& $\mathbf{94.81 \pm 1.83}$
		& $67.13 \pm 2.79$
		& $81.32$ \\
		\bottomrule
	\end{tabular*}
	
	\vspace{3pt}
	\begin{minipage}{\textwidth}
		\footnotesize
		$\checkmark$ denotes confidence filtering of target
		samples; $\times$ denotes using all target samples.
		$^{*}$Proposed configuration.
	\end{minipage}
\end{table*}

\begin{figure}[!t]
	\setlength{\belowcaptionskip}{-0.3cm} 
	\vspace{-0.2cm}
	\centering
	\includegraphics[width=3.0in,height=2.0in]{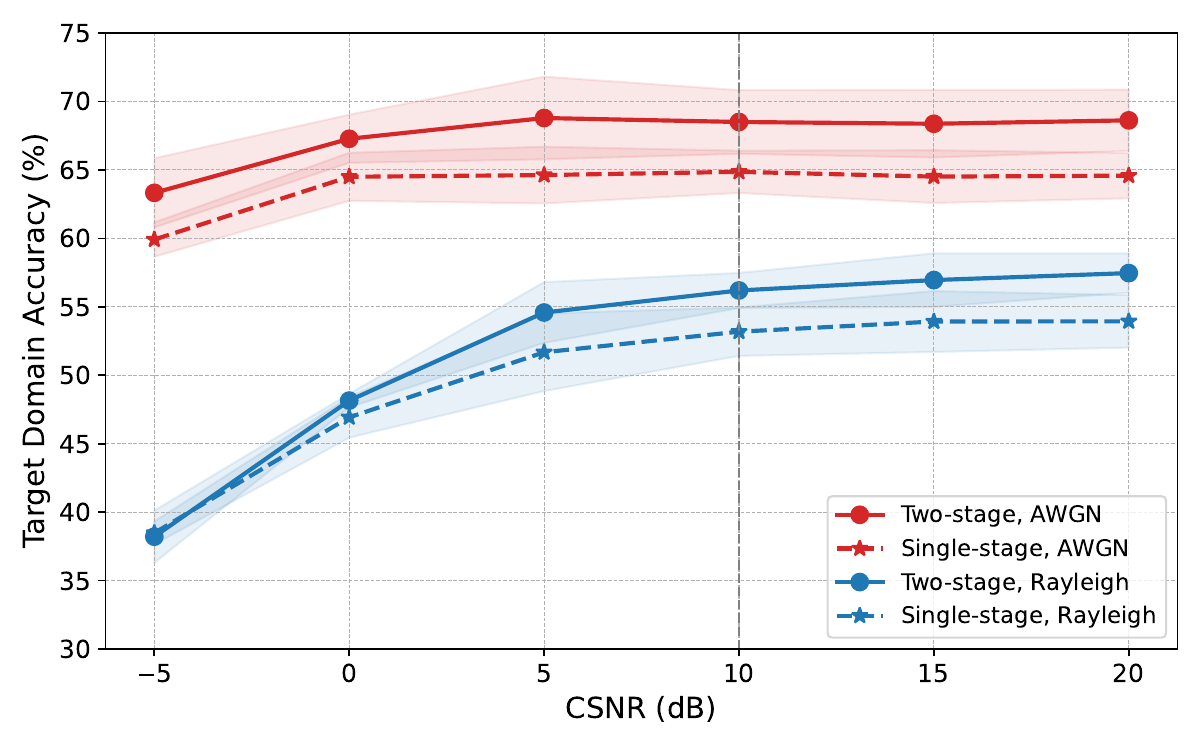}
	\caption{Effectiveness of the proposed two-stage training strategy under the severe A$\rightarrow$S domain shift on PACS.}
	\label{stage}
\end{figure}

\begin{table*}[t]
	\centering
	\renewcommand{\arraystretch}{1.3}
	\setlength{\tabcolsep}{3pt}
	\footnotesize
	\caption{Inference measurements, computational complexity,
		and estimated total training time on the Digits dataset.}
	\label{complexity}
	\begin{tabular}{lcccccc}
		\hline
		Method &
		\shortstack{Inference time} &
		FLOPs &
		Parameters &
		\shortstack{Inference complexity} &
		\shortstack{Training complexity} &
		\shortstack{Training time (h)} \\
		\hline
		Proposed
		& 2.89 ms & 58.15M & 0.55M
		& $\mathcal{O}(K_{\mathrm{back}})$
		& $\mathcal{O}(BK_{\mathrm{back}}+M^2m)$
		& 0.384 \\
		
		Deep JSCC
		& 2.89 ms & 58.15M & 0.55M
		& $\mathcal{O}(K_{\mathrm{back}})$
		& $\mathcal{O}(BK_{\mathrm{back}})$
		& 0.104 \\
		
		KJDM
		& 2.89 ms & 58.15M & 0.55M
		& $\mathcal{O}(K_{\mathrm{back}})$
		& $\mathcal{O}(BK_{\mathrm{back}}+M^2m+M^3)$
		& 0.568 \\
		
		MDAN
		& 7.71 ms & 1.21G & 5.01M
		& $\mathcal{O}(K_{\mathrm{back}}+K_G)$
		& \begin{tabular}[c]{@{}l@{}}
			$\mathcal{O}(B(K_G+K_D+K_{\mathrm{back}}))$
		\end{tabular}
		& $2.517$ \\
		\hline
	\end{tabular}
	\vspace{2pt}
\end{table*}

Table~\ref{tab:mmd_fisher} compares class-conditional
domain discrepancy and target class separability.
The method using only source-domain data exhibits the highest conditional MMD$^2$ and the lowest Fisher ratio, indicating significant domain discrepancy and limited separability of the target classes.
CDA reduces the conditional MMD$^2$ and increases the Fisher ratio.
With the introduction of PCL, the conditional MMD$^2$ is further reduced to 0.0129, while the Fisher ratio improves to 3.510. Thus, using CDA alone significantly improves alignment, and the addition of PCL further enhances the separability of the target classes.

To verify the effectiveness of the proposed two-stage training strategy, Fig. \ref{stage} compares it with a single-stage strategy at AWGN and Rayleigh fading channels under severe domain shift (A$\rightarrow$S). In the single-stage strategy, the complete objective in (\ref{TotalLoss}) is directly applied from the beginning of training. The two-stage strategy consistently achieves higher target-domain classification accuracy than single-stage training over the entire test CSNR range. These results indicate that directly introducing pseudo-label-dependent losses at the beginning of training may cause unreliable predictions, whereas the proposed two-stage strategy provides a more reliable initialization for pseudo-label generation and consequently improves domain adaptation performance. Table \ref{tab:pseudo_label_quality} evaluates pseudo-label quality under the severe A$\rightarrow$S domain shift in PACS. Compared with single-stage training, the two-stage strategy improves accuracy of pseudo-label from 65.29\% to 68.67\% over the AWGN channel and from 54.82\% to 58.97\% over the Rayleigh channel. In addition, the coverage of pseudo-labels has also improved. These results indicate that source-only pretraining before domain adaptation improves pseudo-label reliability without reducing the proportion of selected target samples.

Table~\ref{tab:pacs_confidence_filter_ablation} shows that using all target samples for CDA and filtering them for PCL achieves the highest mean accuracy on three transfer tasks. CDA needs all target samples to cover the target-domain distribution. Filtering PCL samples improves accuracy, supporting the use of high-confidence pseudo labels to construct more reliable contrastive pairs.

\subsection{Complexity Analysis}
Deep JSCC, DANN JSCC, KJDM, and the proposed method share
the same task-oriented JSCC backbone. It consists of a
convolutional feature extractor and linear encoder,
decoder, and classifier heads. We define its per-sample
forward computational cost as
\begin{equation}
	K_{\mathrm{back}}
	=
	\sum_{\ell=1}^{L_f}
	k_\ell^2 C_\ell^{\mathrm{in}} C_\ell^{\mathrm{out}}
	H_\ell W_\ell
	+
	\sum_{j=1}^{L_{\mathrm{fc}}} d_{j-1}d_j,
\end{equation}
where $k_\ell$ is the convolutional kernel size,
$H_\ell$ and $W_\ell$ are the output spatial dimensions,
and $d_{j-1}$ and $d_j$ are the input and output widths
of each linear layer.

Table~\ref{complexity} summarizes inference and training
complexity. Let $B$ and $M$ denote the batchsize in a JSCC training step and the number used in the corresponding contrastive or kernel loss, respectively. The domain discriminators are used during training and discarded at inference. With fixed class and discriminator widths, CDA adds linear batch cost, while PCL computes pairwise similarities. The KJDM method additionally requires dense matrix operations. 

During the inference phase, the adaptation modules are removed, thereby preserving the standard Deep JSCC architecture. In contrast, MDAN needs a generator with cost $K_G$ in inference. 

\section{Conclusion}
This paper studied single-source unsupervised domain adaptation for task-oriented Deep JSCC. We examined how wireless transmission affects domain and class information at the receiver, and established a classification risk bound based on class-conditional domain information. The analysis motivated a joint adaptation method combining class-wise adversarial alignment and pseudo-label supervised contrastive learning on received features. We also related the training loss to the corresponding information objective and analyzed the effect caused by pseudo labels and confidence filtering. Experiments on SVHN$\rightarrow$MNIST and PACS under AWGN and Rayleigh fading channels showed improved target-domain classification. The ablation results confirmed the complementary roles of domain alignment and contrastive learning, while evaluations of confidence filtering and two-stage training highlighted the importance of reliable pseudo labels. Since the adaptation modules are used only during training, the method incurs no additional overhead during the inference phase.

\bibliographystyle{IEEEtran}
\bibliography{IEEEabrv,citation,theory_refs,CCDAJSCC_new_refs}

\vfill

\end{document}